\documentclass[runningheads]{llncs}

\usepackage{color}
\usepackage{graphicx}
\usepackage{amsmath}
\usepackage{subfigure}
\usepackage[dvipsnames]{xcolor}
\usepackage{changepage}

\DeclareMathOperator*{\argmin}{\emph{arg\,min}}

\begin{document}
\title{\vspace{-1cm}RAPPER: {\color{blue}Ra}nsomware {\color{blue}P}revention via {\color{blue}Per}formance Counters\vspace{-0.5cm}}
%
%
%


\author{Manaar Alam\inst{1} \and
Sayan Sinha\inst{1} \and
Sarani Bhattacharya\inst{1} \and
Swastika Dutta\inst{1} \and
Debdeep Mukhopadhyay\inst{1} \and
Anupam Chattopadhyay\inst{2}}

\authorrunning{M. Alam et al.}

\institute{Indian Institute of Technology Kharagpur, India \and Nanyang Technological University, Singapore\\
\email{\{alam.manaar, sayan.sinha, sarani.bhattacharya, swastika, debdeep\}@iitkgp.ac.in, anupam@ntu.edu.sg}\\}

\maketitle              
\vspace{-0.75cm}
\begin{abstract}
Ransomware can produce direct and controllable economic loss, which makes it one of the 
most prominent threats in cyber security. As per the latest statistics, more than half 
of malwares reported in Q1 of 2017 are ransomwares and there is a potent threat of a 
novice cybercriminals accessing ransomware-as-a-service. 
The concept of public-key based data kidnapping and subsequent 
extortion was introduced in 1996. Since then, variants of ransomware 
emerged with different cryptosystems and larger key sizes, the underlying techniques remained same.
Though there are works in literature which proposes a generic framework to detect 
the crypto ransomwares, we present a two step unsupervised detection tool which when suspects a process activity to be malicious, issues an alarm for further analysis to be carried in the second step and detects it with minimal traces.
The two step detection framework- RAPPER uses Artificial Neural Network and Fast Fourier Transformation to develop a highly accurate, fast and reliable solution to ransomware detection using minimal trace points. We also introduce a special detection module for successful identification of disk encryption processes from potential ransomware operations, both having similar characteristics but with different objective. We provide a comprehensive solution to tackle almost all scenarios (standard benchmark, disk encryption and regular high computational processes) pertaining to the crypto ransomwares in light of software security. 
\vspace{-0.3cm}
\end{abstract}

\vspace{-1cm}
\section{Introduction}
\vspace{-0.3cm}
If your organization has not been hit by ransomwares yet, there are chances that it will soon be. The number of medium to large-scale enterprise falling prey to ransom payment and extortion of their private databases have increased manifold. These malicious executables infect victim machines and demand a ransom amount after encrypting files and documents of the device. In May 2017, WannaCry ransomware has affected approximately 400,000 machines across 150 countries~\cite{wannacry_stat}. Identification, blocking of these ransomwares at the earliest along with recovering contents of already encrypted files is already an open challenge.

Hardware Performance Counters (HPCs) were first introduced for checking the static and dynamic integrity of programs to detect any malicious modifications to them as discussed in~\cite{malone2011hardware}. While in~\cite{demme2013feasibility} performance counters are used to build a malware detector in hardware. Detecting malware which modifies the kernel control flow has been targeted in~\cite{karri:13},~\cite{wang2016reusing}. These papers use performance counters to monitor system calls for identifying the vulnerabilities. However, detection of ransomwares through the HPCs, to the best of our knowledge, has not been attempted so far. Though the underlying technique is similar~\cite{simha:14}, ransomware detection requires far more accuracy and faster response time to limit the damage.

A range of ransomwares were studied in~\cite{kharraz_2015_dimva}, which identified 15 different ransomware families. It is suggested that despite advancing encryption systems, the prominent ransomwares leave a trait in the access of I/O and file-systems. Accordingly, Kharraz \emph{et al.}~\cite{kharraz_2016_usenix_automate} proposed a technique of correlating high file system activities with the intrusion of ransomwares, which, however, is susceptible to false positives and also can be defeated with a slow encryption process. Moreover, the technique requires modification in Operating System kernel, which may not be practical in many real-life scenarios. In a recent work, Kiraz \emph{et al.}~\cite{kiraz_2017_eprint_arith} presented a method, where large integer multiplication blocks are identified within an execution. Since public-key cryptosystems rely on large integer multiplications, it can detect the threat at an early stage. Similar approaches for detection of symmetric-key cryptographic primitives via data flow graph isomorphism~\cite{Lestringant_2015_asiaccs_isomorph} or by identifying characteristics of a cipher in a binary code~\cite{grobert_2011_book_binary} are also presented. In this paper, neither we target a specific family of ransomwares nor the properties corresponding to a particular cipher implementation used by a ransomware. Instead, we develop a generic anomaly-based approach based on the HPC statistics.

A major difficulty in any ransomware detection approach is to differentiate between a benign disk encryption process and a ransomware executable. While both of them serve the same objective, but the latter being unintended and malicious one. Most of the popular disk encryption applications require administrative privilege and use similar algorithms in their encryption operation. In this paper, we utilize this fact and try to address the harder problem to differentiate between a disk encryption program and a ransomware not only by checking the privilege of a program but also by observing its behavior in terms of hardware performance counters. We also present a simple yet efficient method to recover the files encrypted by a ransomware before its detection by utilizing the Linux file locking mechanism using \texttt{mlock()} system call.

\vspace{-0.5cm}
\subsection*{Motivation and Contribution}
\vspace{-0.1cm}
The primary contributions of this paper are listed below:
{\small 
\begin{itemize}
\item Main objective of RAPPER is to learn the behavior of system under observation with performance event statistics obtained from HPCs. Unlike other works in literature, which save the templates of malicious processes and matches it on its occurrence, here we allow our tool to learn the normal operating behavior of the system. The time-series data as observed from a selected cluster of HPC events is fed to an Artificial Neural Network to learn the specific characteristics of the data.
\item Any deviation from this normal behavior as learned by the Autoencoder is considered as a suspect to RAPPER. We observed that the performance statistics of the system in presence of ransomware are significantly dissimilar from the normal system behavior because of repeated encryption process.
\item In the course of our study, we came across a benchmark application, though benign in nature, raised an alarm. This is a typical example of a benign process, which due to its high computational overhead differs significantly from normal system behavior, thus raising a false alarm. 
\item On such an alarm, we transform the time series into frequency domain using Fast Fourier Transformation (FFT) and understand the repeatability of data with the help of a second autoencoder.
\item We also explore the performance of RAPPER in the presence of disk encryption programs, which are benign in nature having similar behavior as ransomwares, and devised a correlation based approach to differentiate between these two processes. 
\end{itemize}}
\vspace{-0.3cm}
RAPPER is a lightweight tool, which does not require any hardware or kernel modification, thereby making it practical to use in almost every environment.

\vspace{-0.5cm}
\section{Anomaly Detection by Analysing the System Behavior}
\vspace{-0.3cm}
In this section, we first analyze the \emph{normal behavior}\footnote{We term regular execution pattern of an uncompromised system as \emph{normal behavior}, which captures all the daily operations of benign executables assuming none of them are ransomwares. We do not consider the execution patterns of high-computation programs within normal behavior, as ransomwares, which are also high-computation programs, may be missed by the detection framework by doing so.} of a system by monitoring some appropriately selected hardware performance counter events in parallel. We then present a notion of anomalous activity in the system and demonstrate a detailed methodology for detecting those anomalies by using an \emph{Autoencoder}.

\vspace{-0.3cm}
\subsection{Observing the System Behavior using HPCs}\label{sec:hpc}
The Hardware Performance Counters (HPCs) are a set of special purpose registers built into modern processors to dynamically observe the hardware related activities in a system. There are some recent works~\cite{karri:13},~\cite{wang2016reusing} which use these HPCs to detect malicious programs targeted for a particular system. The HPCs can be monitored dynamically using the well-known \texttt{perf} tool, available in Linux kernels 2.6.31 and above. One interesting property of the perf tool is that a user can observe the performance counters associated with a system with some time interval, thereby giving the benefit of observing the system behavior continuously in a succession of time. The command to monitor a particular HPC event for a specific executable in such way is as follows:

\vspace{-0.3cm}
\begin{center}
	\texttt{perf stat -e <event\_name> -I <time\_interval> <executable\_name>}
\end{center}
\vspace{-0.3cm}

The range of HPC events those can be monitored using the perf tool is more than 1000. However, in most of the Linux based systems, the perf tool is limited to observing a maximum of 6 to 8 hardware events in parallel depending on the processor type. Moreover, some of the events are not even supported by all the processors. Our objective, in this work, is to detect the presence of ransomwares, which mainly contain an encryption program, typically involving both symmetric and asymmetric key encryptions. Hence, we selected the hardware events which are more likely to change because of the encryptions and are supported in most of the processors. The hardware events selected for our study are \texttt{instruction}, \texttt{cache-references}, \texttt{cache-misses}, \texttt{branches}, and \texttt{branch-misses}. The events are self-explanatory by their names. Generally, the symmetric encryption affects the cache based events while the asymmetric encryptions affect the instruction and branching events.

In order to represent the prototype of normal system behavior, we designed a \emph{watchdog program} executable and collected the \texttt{perf stat} values with \texttt{10ms} time interval\footnote{The minimum interval of time after which the perf tool is allowed to sample a data point is 10ms. We have selected the minimum time-interval to sample data as fast as allowed, to enhance the mechanism in terms of detection time.} for that executable. We collected these values at the different point of time in the target system and created a dataset of regular observation. The effects of all the other processes including the ransomwares running in the system will have an impact on the performance counters values. We articulate that any behavior which is not close to this dataset is unusual activity, but may not be a malicious one.

\begin{figure}[!t]
	\begin{adjustwidth}{-3cm}{-3cm}
	\centering
	\subfigure[\textbf{\# Branch Instructions}]{
		\includegraphics[width=0.35\textwidth]{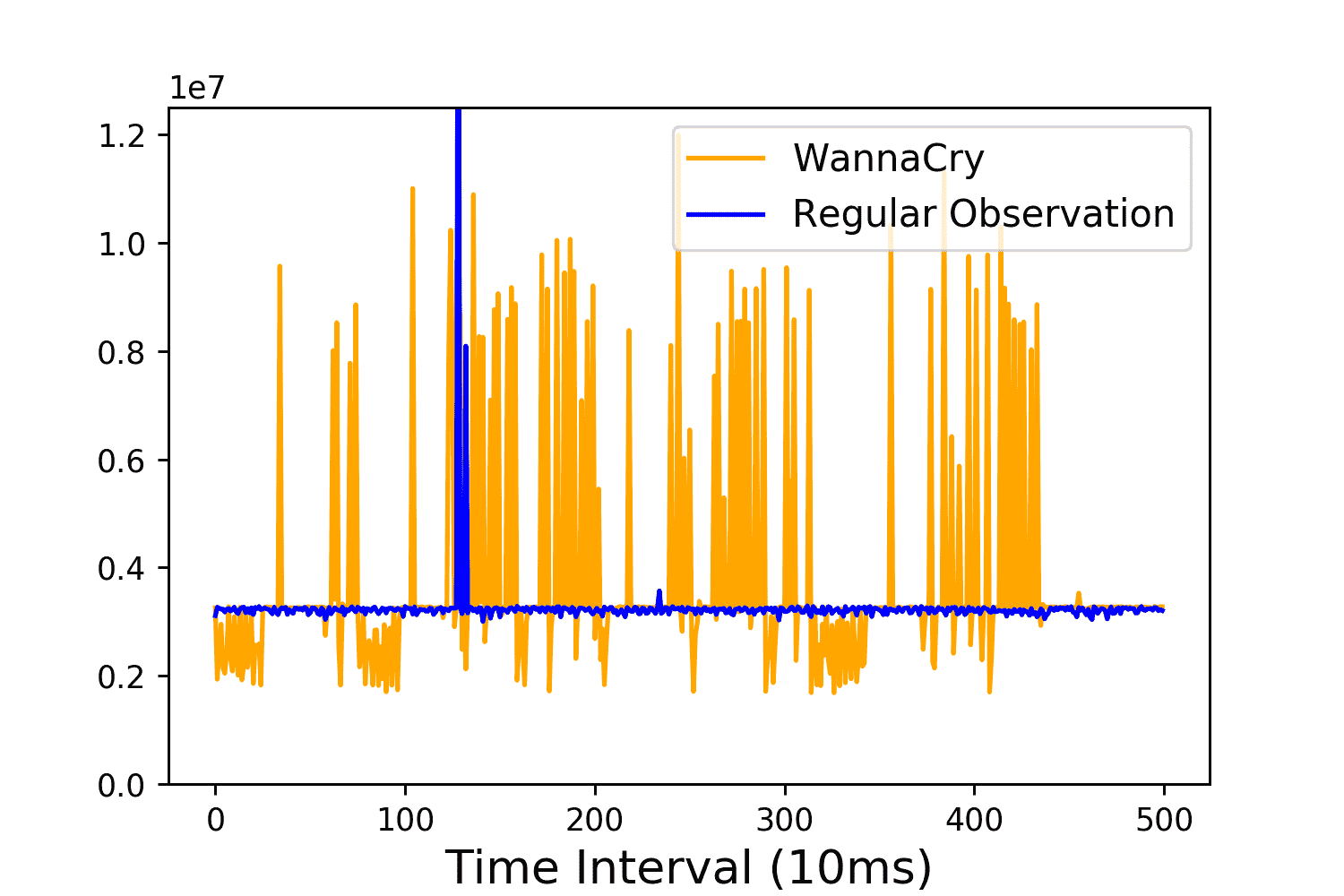}
		\label{fig:observe_branch}}
	\subfigure[\textbf{\# Branch Misses}]{
		\includegraphics[width=0.35\textwidth]{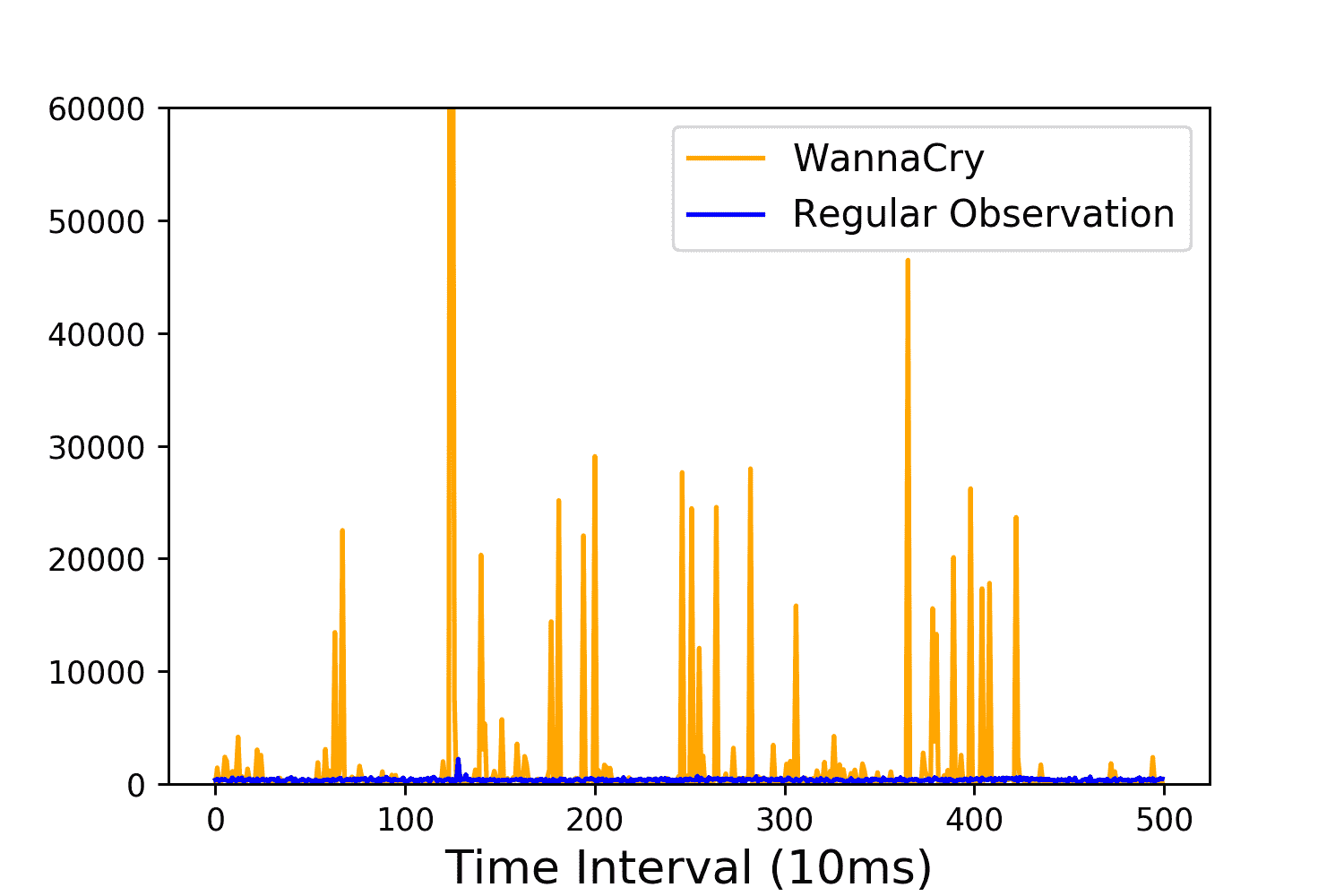}
		\label{fig:observe_br_miss}}
	\subfigure[\textbf{\# Cache Misses}]{
		\includegraphics[width=0.35\textwidth]{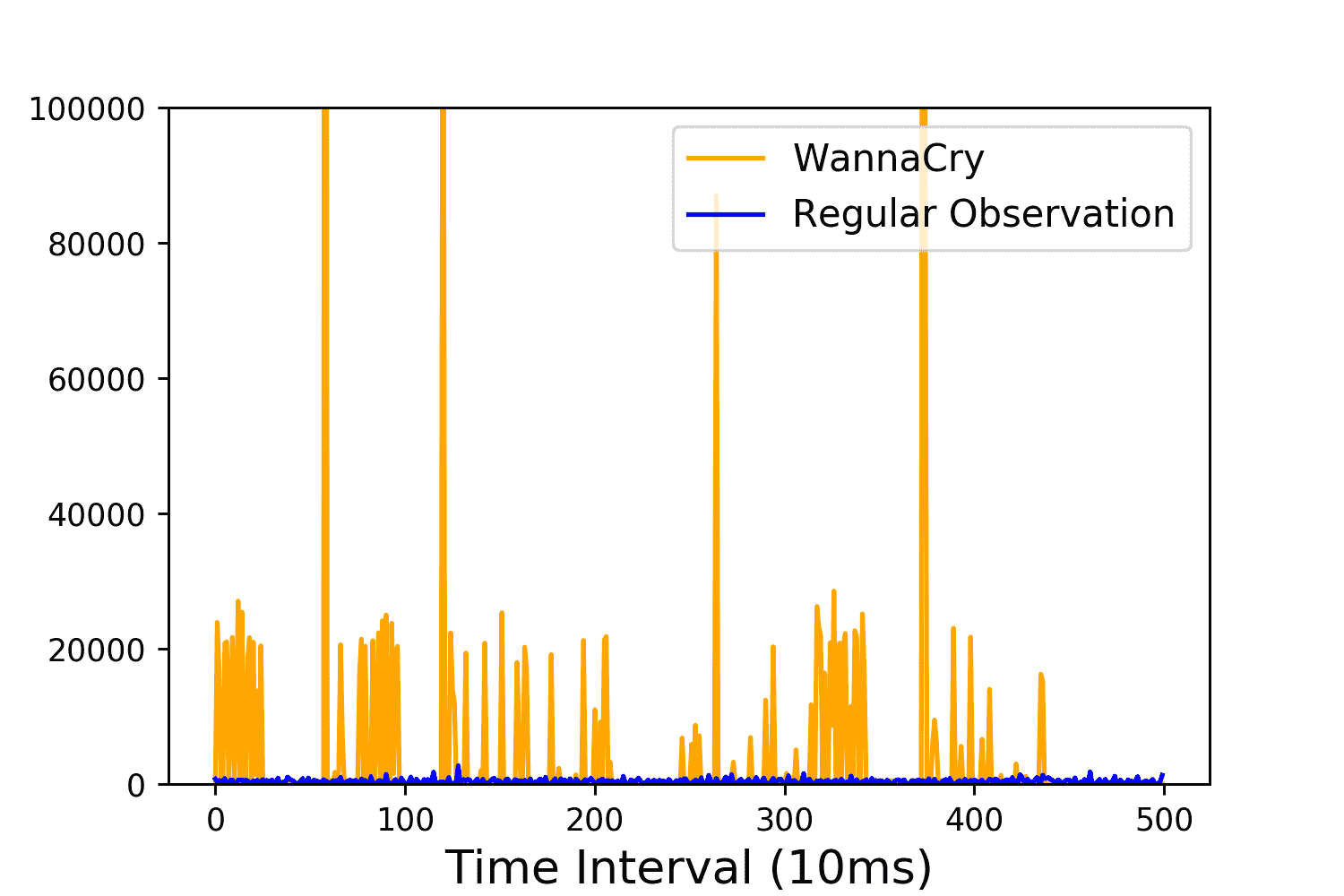}
		\label{fig:observe_ca_miss}}
	\subfigure[\textbf{\# Cache References}]{
		\includegraphics[width=0.35\textwidth]{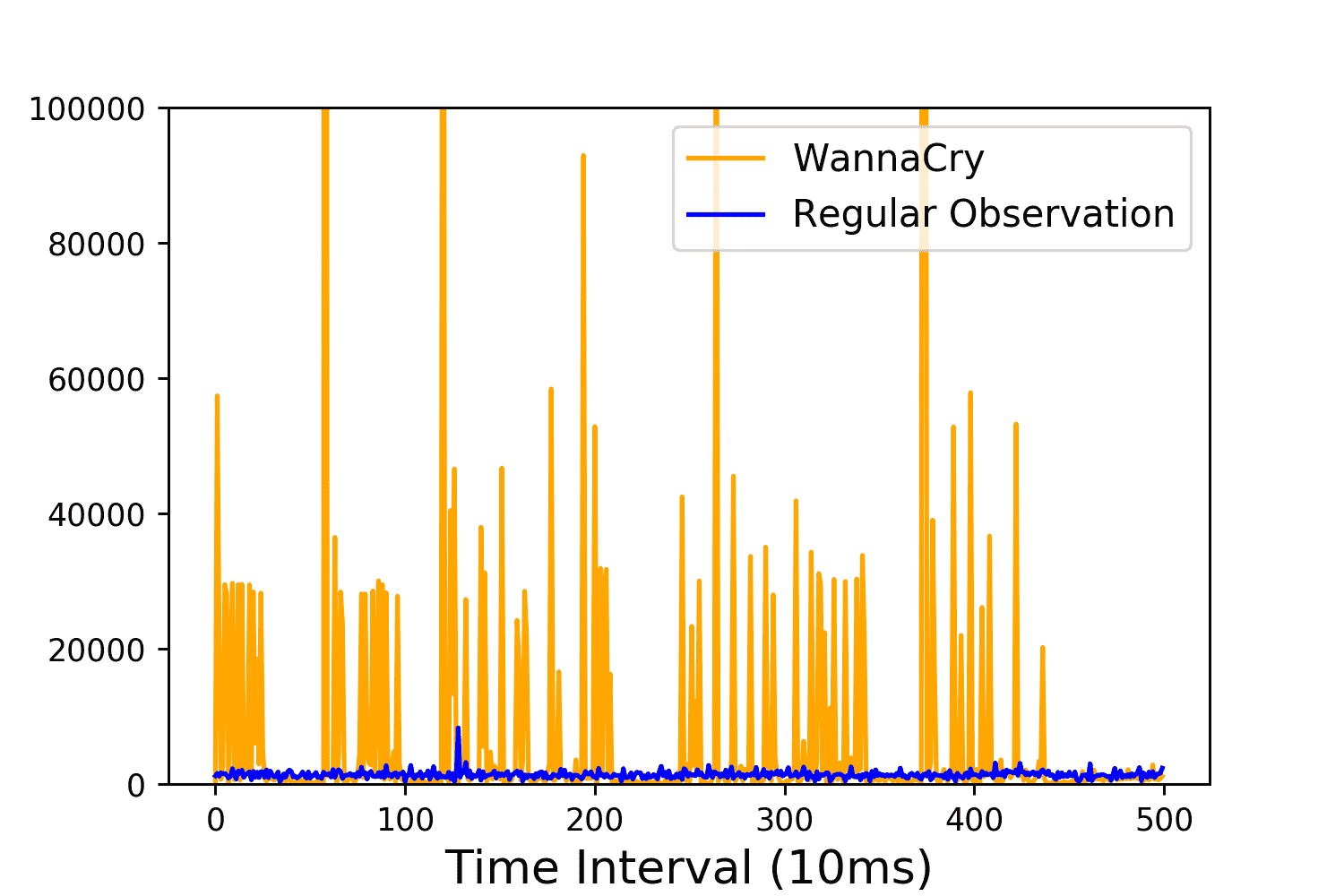}
		\label{fig:observe_cache}}
	\subfigure[\textbf{\# Instructions}]{
		\includegraphics[width=0.35\textwidth]{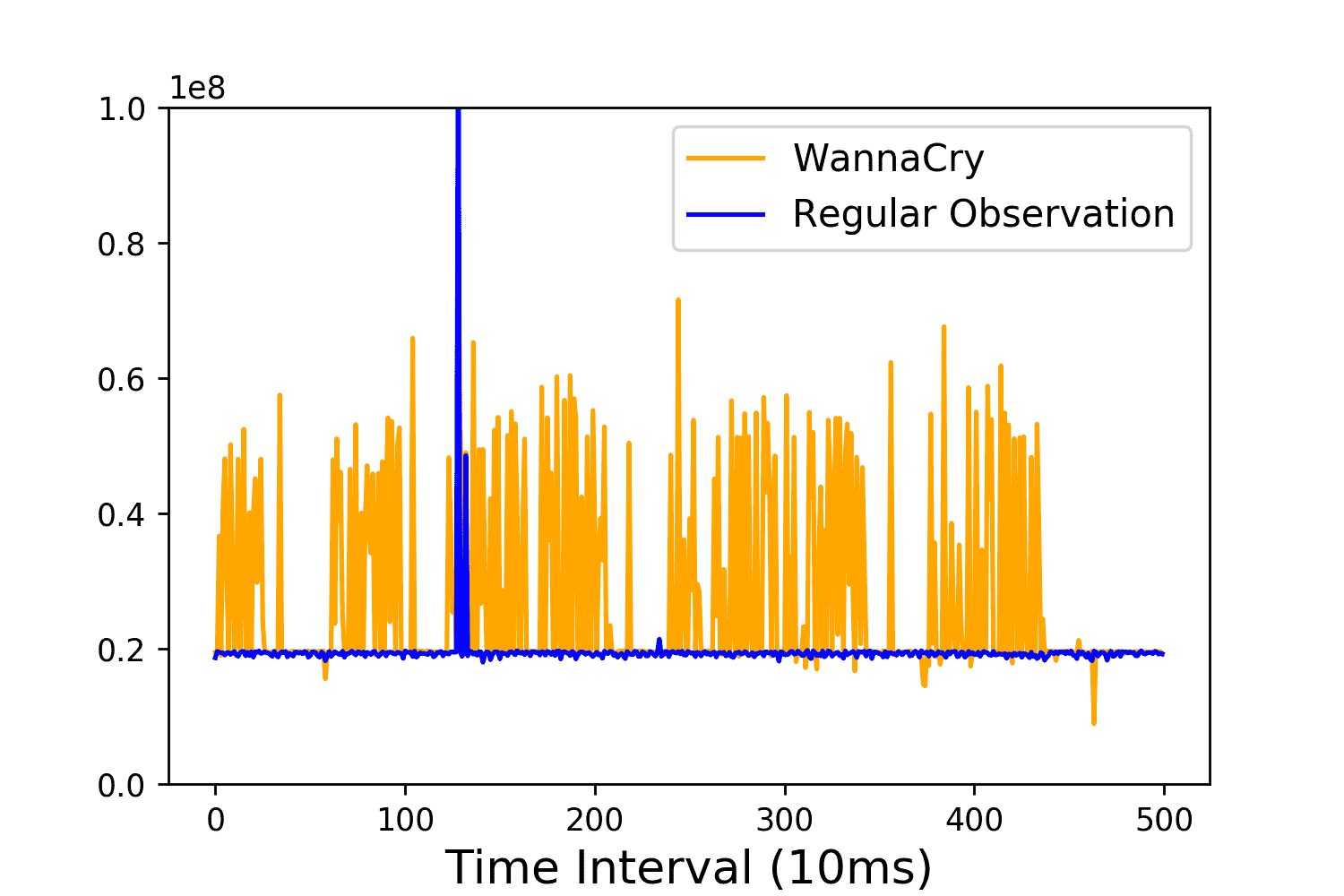}
		\label{fig:observe_ins}}
    \end{adjustwidth}
    \vspace{-0.3cm}
    \caption{Variation of Performance Counter Events in presence of Wannacry Ransomware\label{fig:perf_dev}\vspace{-0.5cm}}
\end{figure}

We show the effect of a Ransomware Program (for example, a WannaCry) on the HPC values in Fig.~\ref{fig:perf_dev}. The \textcolor{blue}{blue} lines in Figure represent the effect of normal system programs on the watchdog executable for different HPCs, whereas the \textcolor{orange}{orange} lines show the effect of WannaCry ransomware.

\begin{figure}[!t]
	\begin{adjustwidth}{-4cm}{-4cm}
		\centering
		\subfigure[]{
			\includegraphics[width=0.5\textwidth]{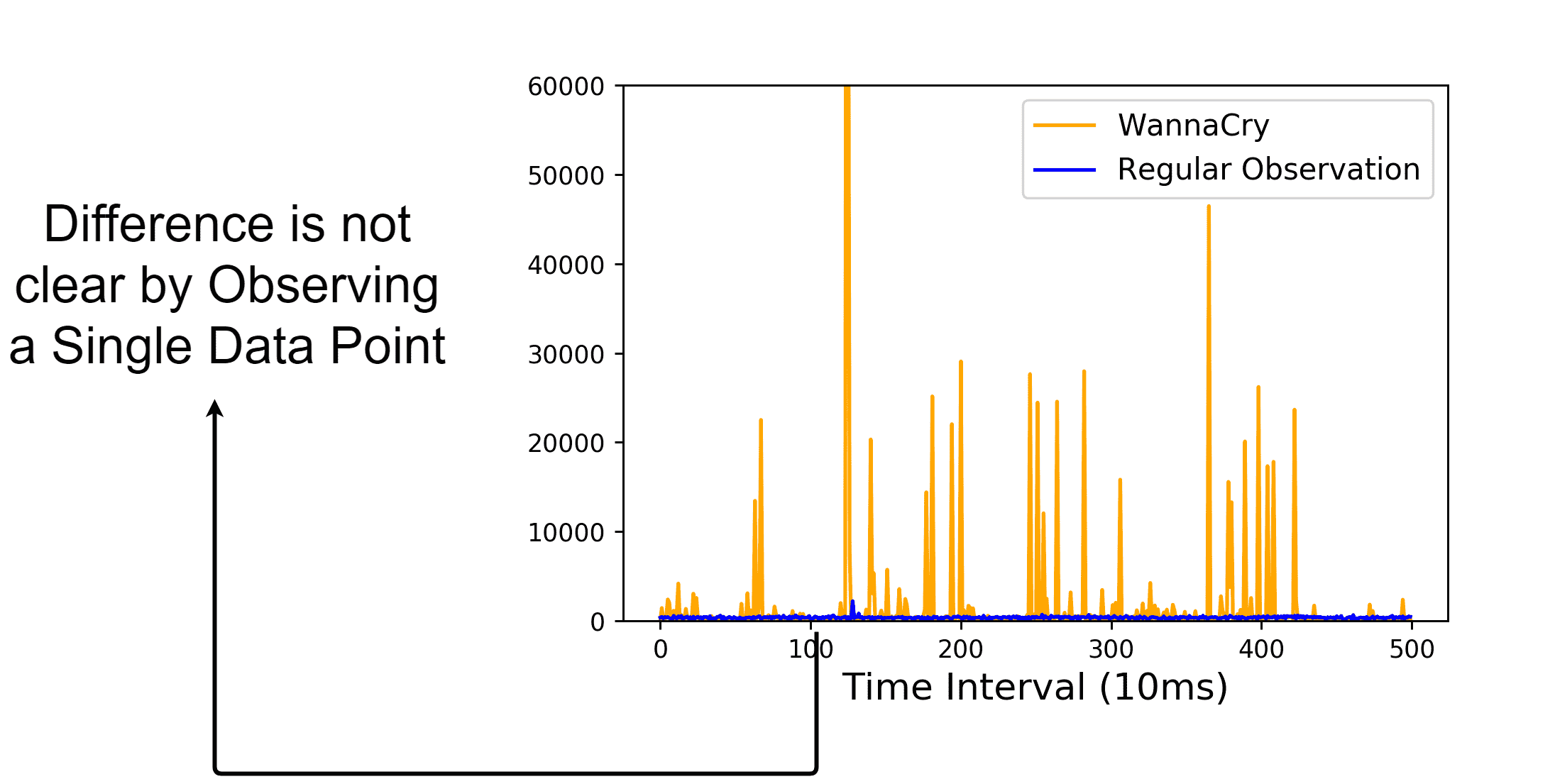}
			\label{fig:branch_miss_no_window}}
		\subfigure[]{
			\includegraphics[width=0.5\textwidth]{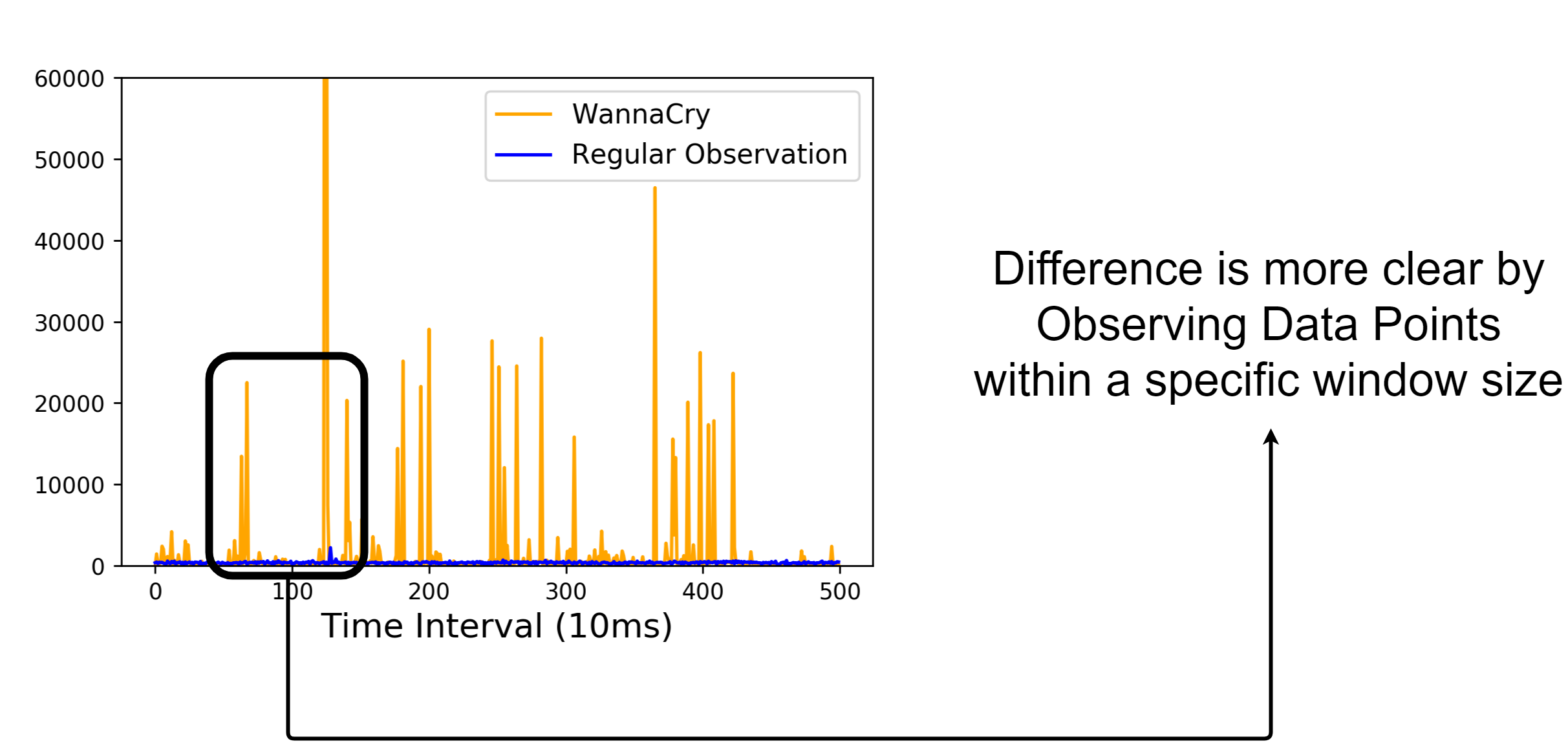}
			\label{fig:branch_miss_with_window}}
	\end{adjustwidth}
	\vspace{-0.3cm}
	\caption{Importance of observing data-points within a window instead of single data-point for decision making considering HPC values for the event \texttt{branch-misses}\label{fig:imp_window}\vspace*{-0.6cm}}
\end{figure}

An important point to be observed is that for a particular interval of time the behavior of WannaCry ransomware does not change much from the normal system behavior. For example, around time interval of 100, as shown in Fig.~\ref{fig:branch_miss_no_window}, the effect of WannaCry on the hardware event branch misses is same as normal system behavior. However, if we consider a window of a specific length, as shown in Fig.~\ref{fig:branch_miss_with_window}, the behavior of WannaCry is more distinguishable from normal observation. So, instead of considering individual points for decision making, we select a window of observations considering each of the five events collectively. Thus, we transform the problem into anomaly detection in multivariate time-series data.

\vspace{-0.5cm}
\subsection{Learning a Time-Series Data Using an Autoencoder}\label{sec:learn_auto}
\vspace{-0.3cm}
In order to present a generalized ransomware detection strategy, we avoid modeling the behavior of ransomwares as there can be a potential new one whose behavior is unknown. Instead, we model the normal system behavior, as we can get a majority of such instances. Another advantage of detecting anomalies by modeling normal behavior is that we do not need the necessity of labeled dataset as any activity with unusual behavior crossing an empirically calculated threshold value can be detected as an anomaly. Thus, we propose an unsupervised approach to detect these anomalies. We have already seen that HPC values observed over the watchdog application is considered as the time-series data. An LSTM (Long-Short-Term-Memory) based autoencoder can efficiently implement the unsupervised anomaly detection for time-series, which we discuss below.

Autoencoder is a Deep Artificial Neural Network used for efficient coding of the input space by unsupervised learning. The primary goal of an autoencoder is to induce a representation for a set of data by learning an approximate identity function, i.e., if the input data is $\mathcal{X}$, the goal of the autoencoder is to learn the function $f$, given by - \hspace{1cm}
	$f : \mathcal{X} \rightarrow \mathcal{X}$

An autoencoder always consists of two mapping, encoding and decoding, which are given as $\phi$ and $\psi$ respectively. \hspace{1cm}
      $\phi : \mathcal{X} \rightarrow \mathcal{F}$, \hspace{0.5cm}
      $\psi : \mathcal{F} \rightarrow \mathcal{X}$

where $\mathcal{F}$ is a vector referring to the decisive intermediate representation learned by the autoencoder, which is used to regenerate the original input data. The error incurred by the autoencoder to regenerate the input from vector $\mathcal{F}$ is termed as \emph{Reconstruction Error}, which is given as: \hspace{0.5cm}
$\mathcal{L} = \| \mathcal{X} - (\psi \circ \phi)\mathcal{X}\|^2$

The goal of the autoencoder is to minimize these reconstruction error for all the input samples, i.e., to find the mappings $\phi$ and $\psi$ such that $\mathcal{L}$ is minimum.

\vspace{-0.3cm}
\begin{equation}
	\label{eq:rec_err}
	\argmin_{\phi, \psi} \mathcal{L} = \argmin_{\phi, \psi} \| \mathcal{X} - (\psi \circ \phi)\mathcal{X}\|^2
\end{equation}
\vspace{-0.4cm}

In our case, the input $\mathcal{X}$ is a multivariate time-series sequence, and the objective is to learn the structure of the sequence. LSTM networks, belonging to a class of Recurrent Neural Network Model, are typically used for modeling sequence data, which efficiently handles the dependencies within the sequence. Hence, we use the LSTM based autoencoder for our detection purpose. The anomaly detector model first takes a multivariate input sequence ($\mathcal{X}$), generates an intermediate feature vector ($\mathcal{F}$) related to the sequence, and then reconstructs the same sequence from the intermediate feature vector. The autoencoder is trained using all the input sequences of regular observation by following the objective function mentioned in Equation~(\ref{eq:rec_err}).

The training dataset is constructed from the observed data for normal system behavior by taking a window of 100 trace points\footnote{The window size of 100 is chosen empirically.} (i.e., a window trace points collected over 1 second, since each interval data is collected after 10ms). We shift the window by one time-interval (i.e., 10ms) repeatedly to consider consecutive 100 sample point for learning. Once the learning of intermediate vector $\mathcal{F}$ is completed, for an anomalous sequence, the autoencoder makes an attempt to reconstruct the original input sequence. Thus, the autoencoder maps it to the normal sequence, based on the intermediate feature vector $\mathcal{F}$. There is an inherent information loss in this process and hence will incur a substantial \emph{reconstruction error}. Next, we quantify the amount of error to be incurred by a process to be termed as an anomaly.

\vspace{-0.5cm}
\subsection{Determining Threshold for the Decision}\label{sec:threshold}
\vspace{-0.3cm}
In order to quantify the threshold for detecting anomalous activities, we calculate the reconstruction error distribution ($\mathcal{R}$) for all the samples in normal behavior. According to the $3\sigma$ rule of thumb, all these values should lie within three standard deviations of the mean. Hence, we set the threshold for reconstruction error ($\mathcal{R}_t$) as below.

\vspace{-0.75cm}
\begin{equation}
\label{eq:three_sigma}
\mathcal{R}_{t} = \mu_{\mathcal{R}} + 3 * \sigma_{\mathcal{R}}
\end{equation}
\vspace{-0.5cm}

where $\mu_{\mathcal{R}}$ and $\sigma_{\mathcal{R}}$ are the mean and standard deviation of distribution $\mathcal{R}$. In our experimental setup $\mathcal{R}_{t}$ came out to be $5.38 \times 10^{-6}$.

\vspace{-0.5cm}
\subsection{Anomalous Behaviors of Ransomwares}
\vspace{-0.3cm}
\begin{figure}[!t]
    \centering
    \subfigure[\textbf{WannaCry}]{
    	\includegraphics[width=0.35\textwidth]{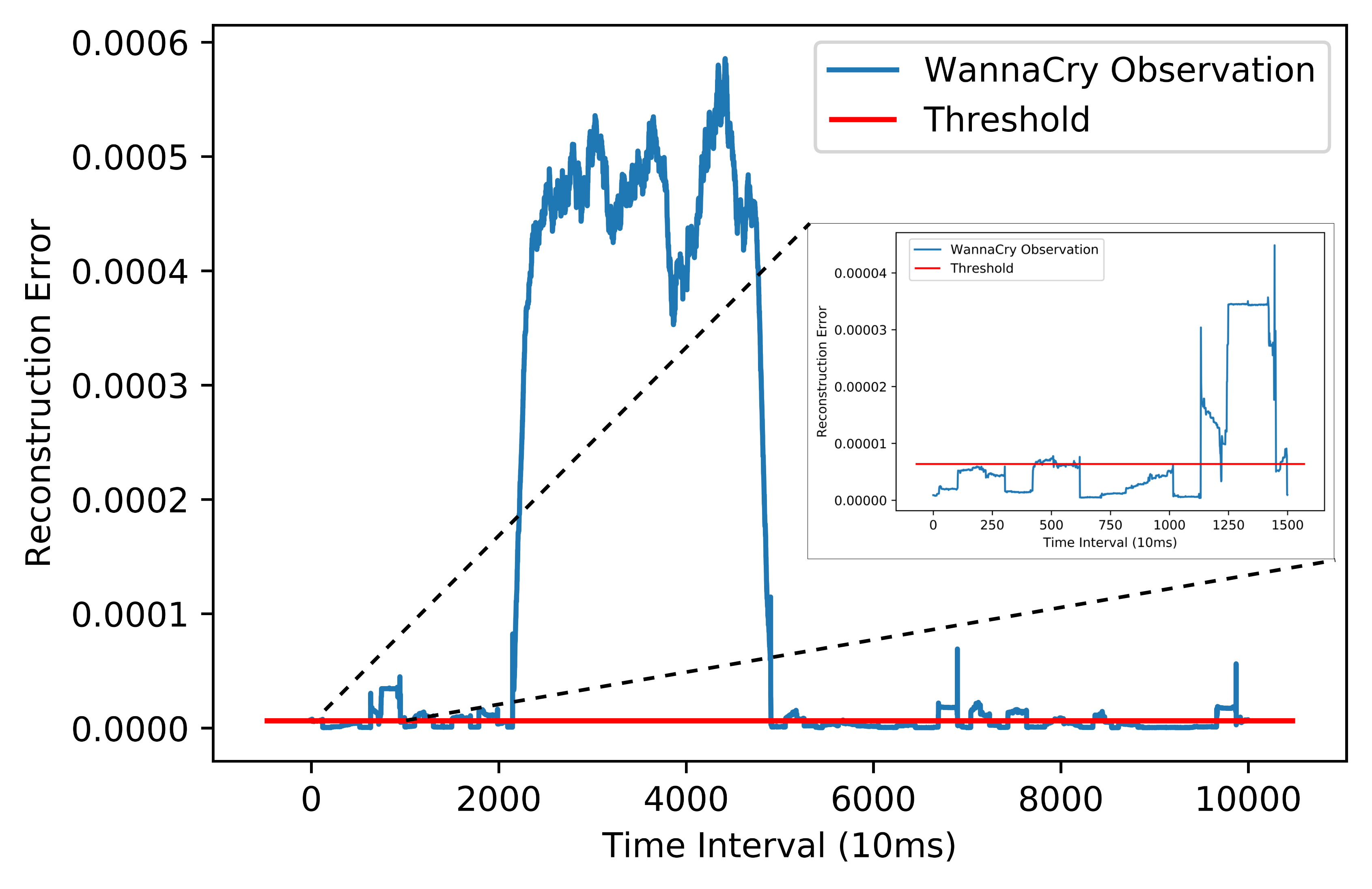}
    	\label{fig:rec_wannacry}}
    \subfigure[\textbf{Vipasana}]{
    	\includegraphics[width=0.35\textwidth]{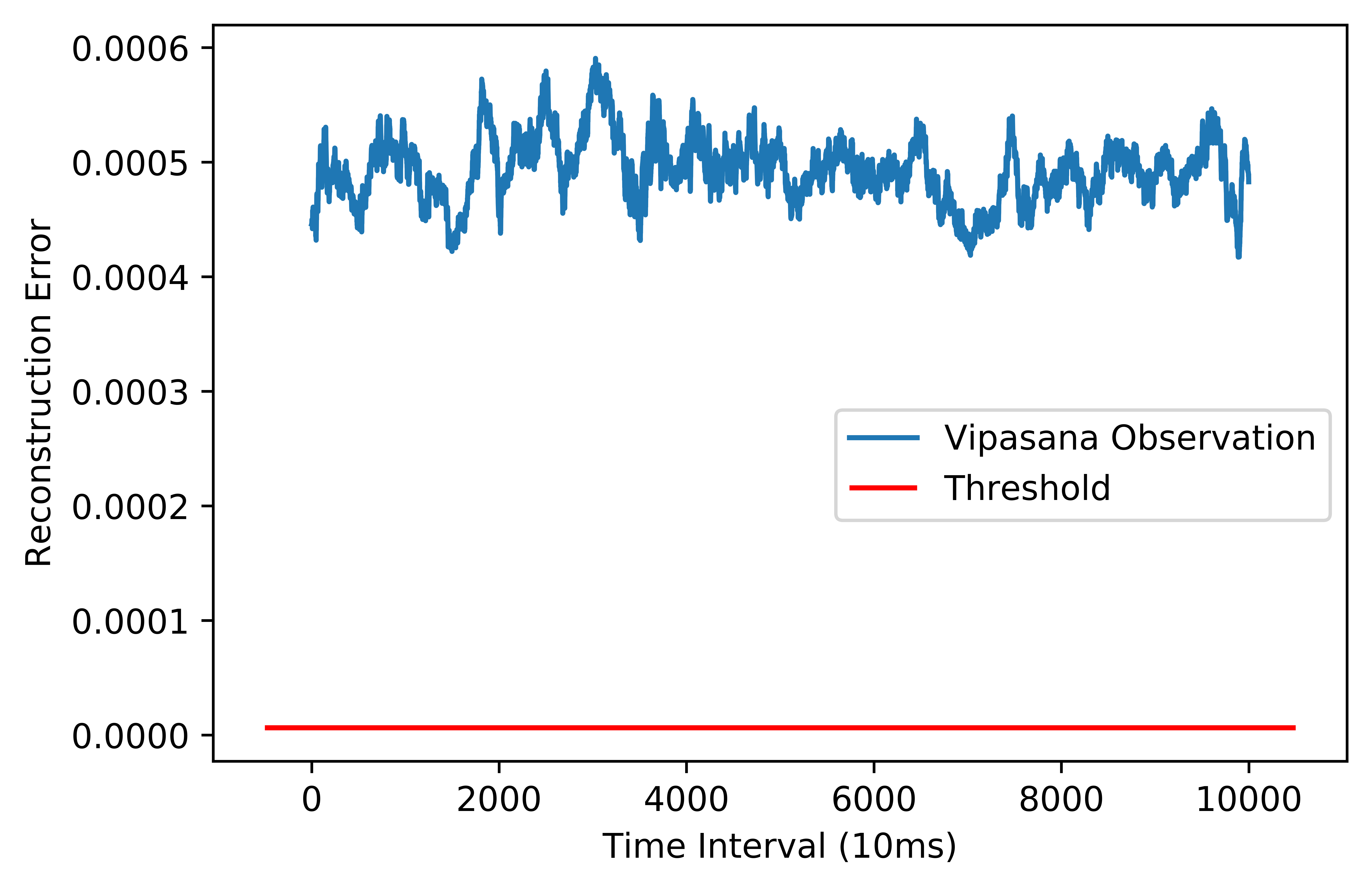}
    	\label{fig:rec_vipasana}}
    \subfigure[\textbf{Locky}]{
    	\includegraphics[width=0.35\textwidth]{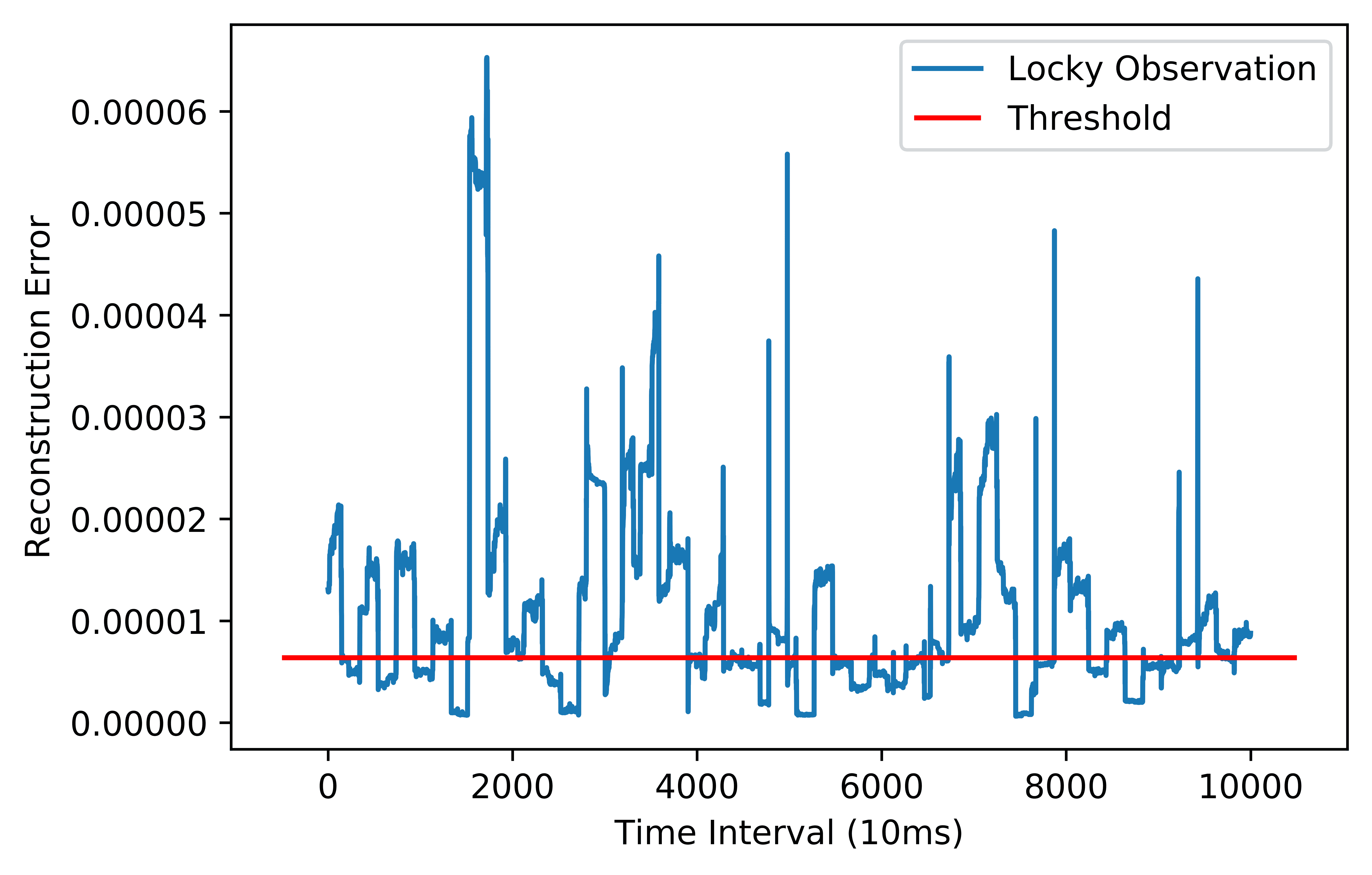}
    	\label{fig:rec_locky}}
    \subfigure[\textbf{Petya}]{
    	\includegraphics[width=0.35\textwidth]{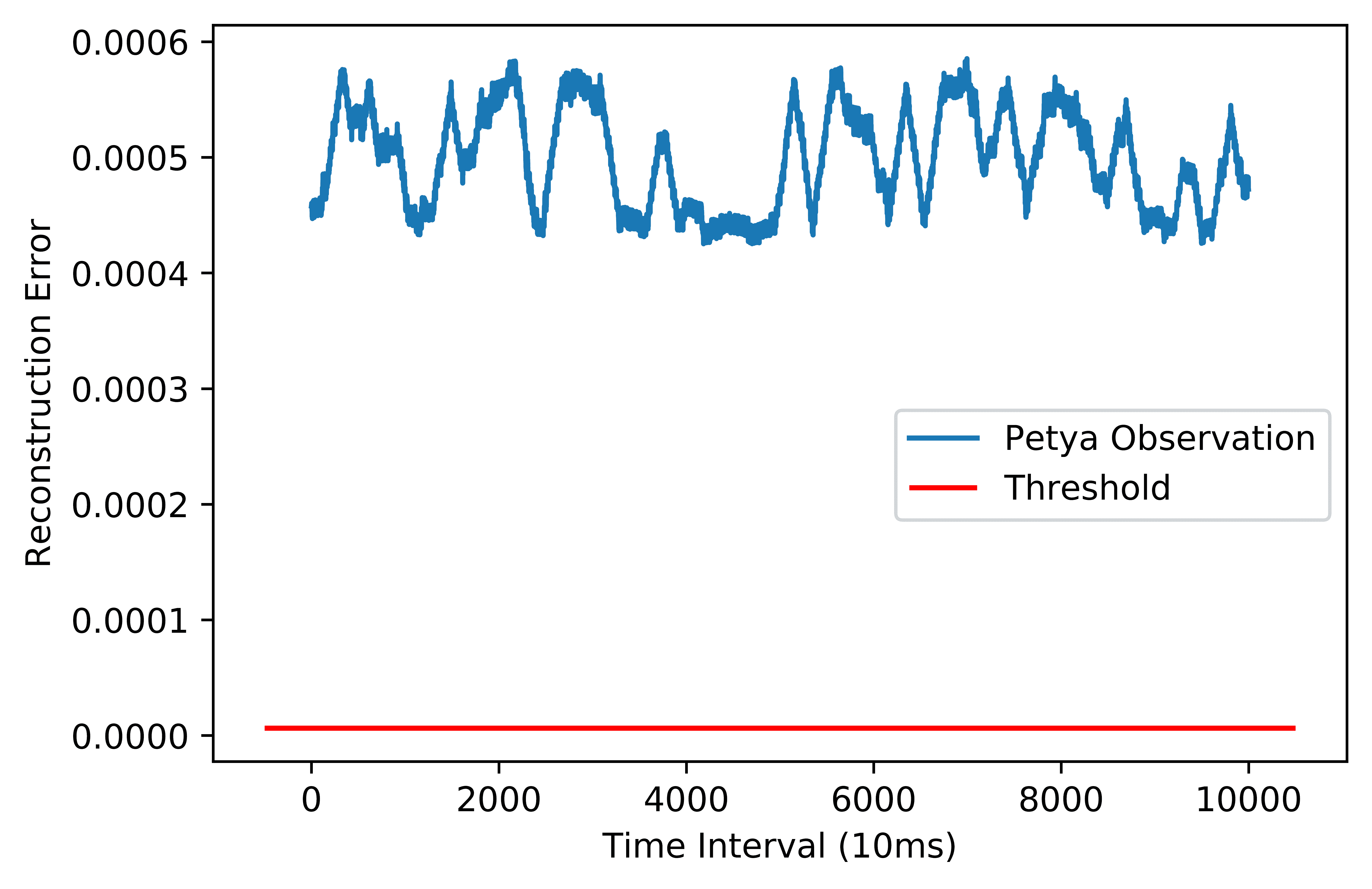}
    	\label{fig:rec_petya}}
	\vspace{-0.3cm}
    \caption{Sequence of Reconstruction Errors for Ransomware in Autoencoder\_1\label{fig:seq_rec_error}\vspace{-0.5cm}}
\end{figure}

In our study, we considered four ransomware programs - namely \emph{WannaCry}, \emph{Vipasana}, \emph{Locky}, and \emph{Petya} to show the impact of selecting the threshold $\mathcal{R}_t$ in detecting them as anomalies. Fig.~\ref{fig:seq_rec_error} shows the sequence of reconstruction errors for these ransomwares. The first point on these plots represents the first window of 100 time-interval (equivalently 1 second). The successive points come after each interval of 10ms as we slide by one time-interval for calculating the next reconstruction error. The \textcolor{blue}{blue} line indicates the reconstruction errors of each window whereas the \textcolor{red}{red} line signifies the threshold $\mathcal{R}_t$ as calculated before.

We can observe from Fig.~\ref{fig:rec_wannacry}\footnote{The embedded image in the box is the zoomed version of the same dataset for the first 1500 window data.}, the execution of WannaCry starts behaving like a regular program (since the reconstruction errors lie well below the threshold value), but the reconstruction error shoots over the threshold at $432^{nd}$ observation. Thus, the WannaCry is detected as anomaly $(1000 + 431*10) = 5310$ ms or $5.31$ seconds after the start of execution. Whereas, from Fig.~\ref{fig:rec_vipasana}, Fig.~\ref{fig:rec_locky}, and Fig.~\ref{fig:rec_petya}, we can observe that the ransomwares Vipasana, Locky, and Petya are detected as an anomaly at the first window itself, i.e., $1$ second after the start of execution. In all these cases there is an extra overhead of time due to the testing time associated with the Autoencoder, which we discuss in Section~\ref{sec:results}.

\vspace{-0.5cm}
\section{How Good is Reconstruction Error as a Decider?}
\vspace{-0.3cm}
In the previous section, we suggested that a threshold as high as $\mathcal{R}_t$ can be used to decide whether a particular process behavior deviates from the normal system behavior significantly. In this section, we explain why a single decision step is not enough to claim that the anomaly observed is from a malicious process.

\vspace{-0.5cm}
\subsection{Understanding the Ambiguity}
\vspace{-0.3cm}
\begin{figure}[!t]
	\begin{adjustwidth}{-3cm}{-3cm}
	\centering
	\subfigure[\textbf{\# Branch Instructions}]{
		\includegraphics[width=0.35\textwidth]{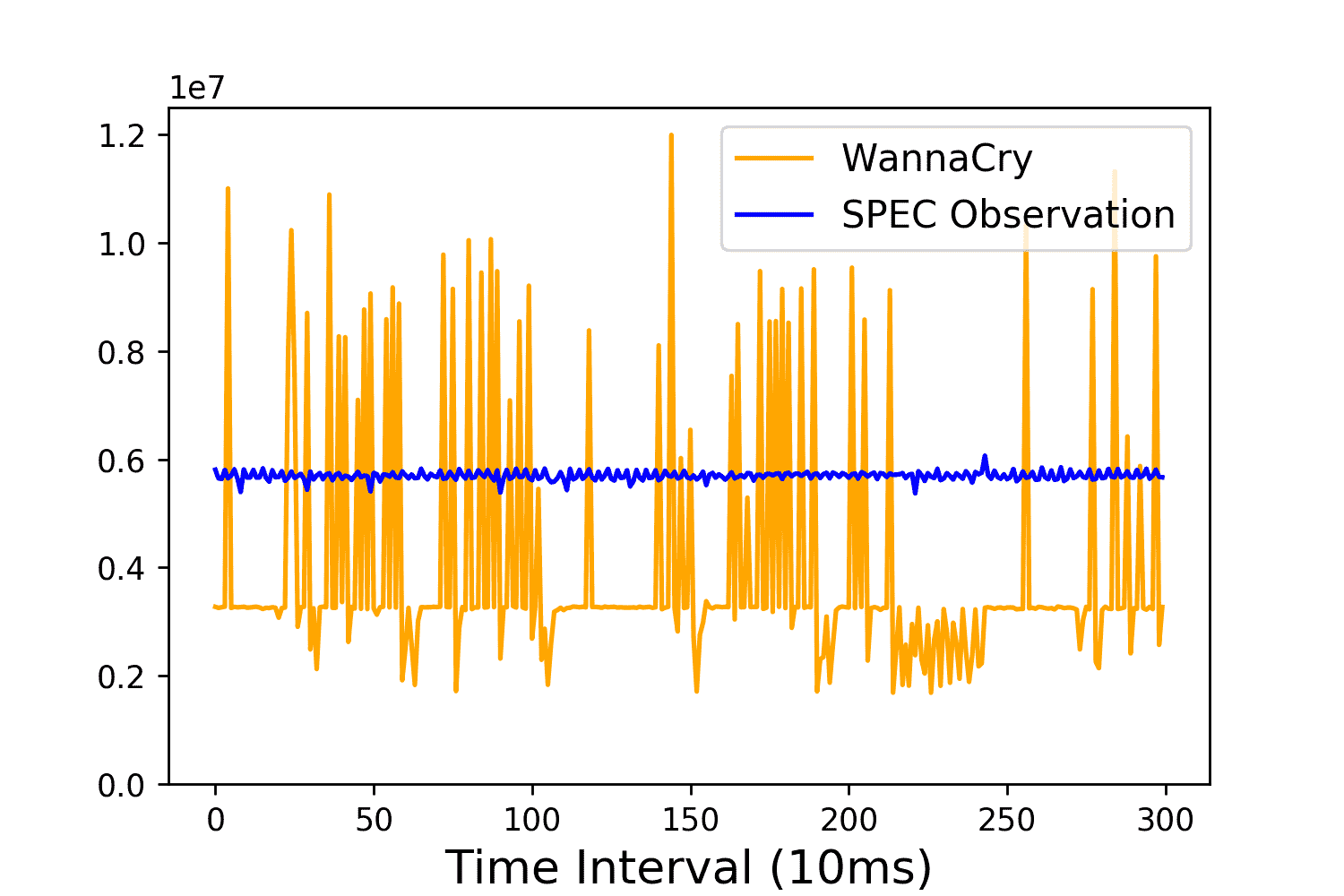}
		\label{fig:com_branch}}
	\subfigure[\textbf{\# Branch Misses}]{
		\includegraphics[width=0.35\textwidth]{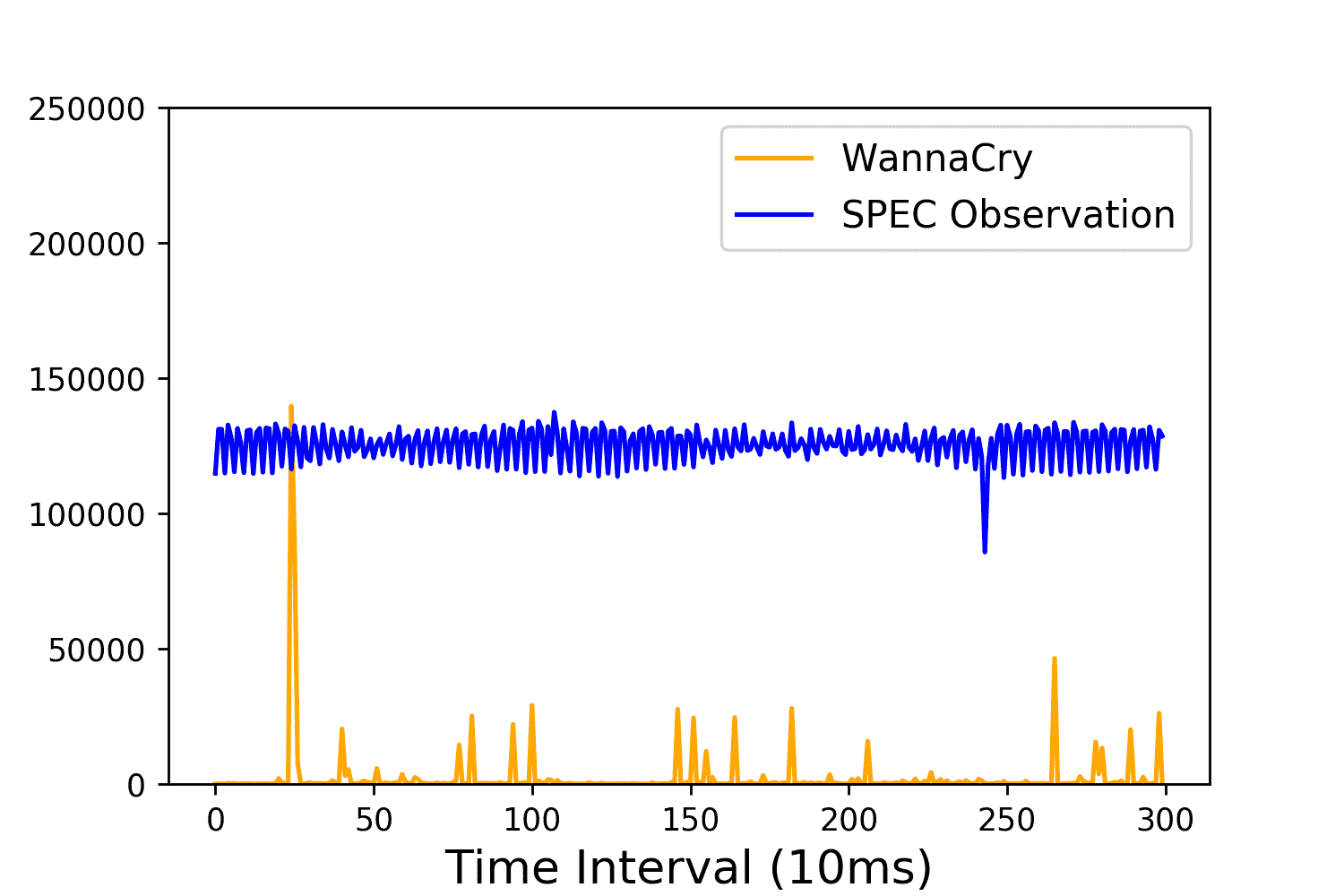}
		\label{fig:com_br_miss}}
	\subfigure[\textbf{\# Cache Misses}]{
		\includegraphics[width=0.35\textwidth]{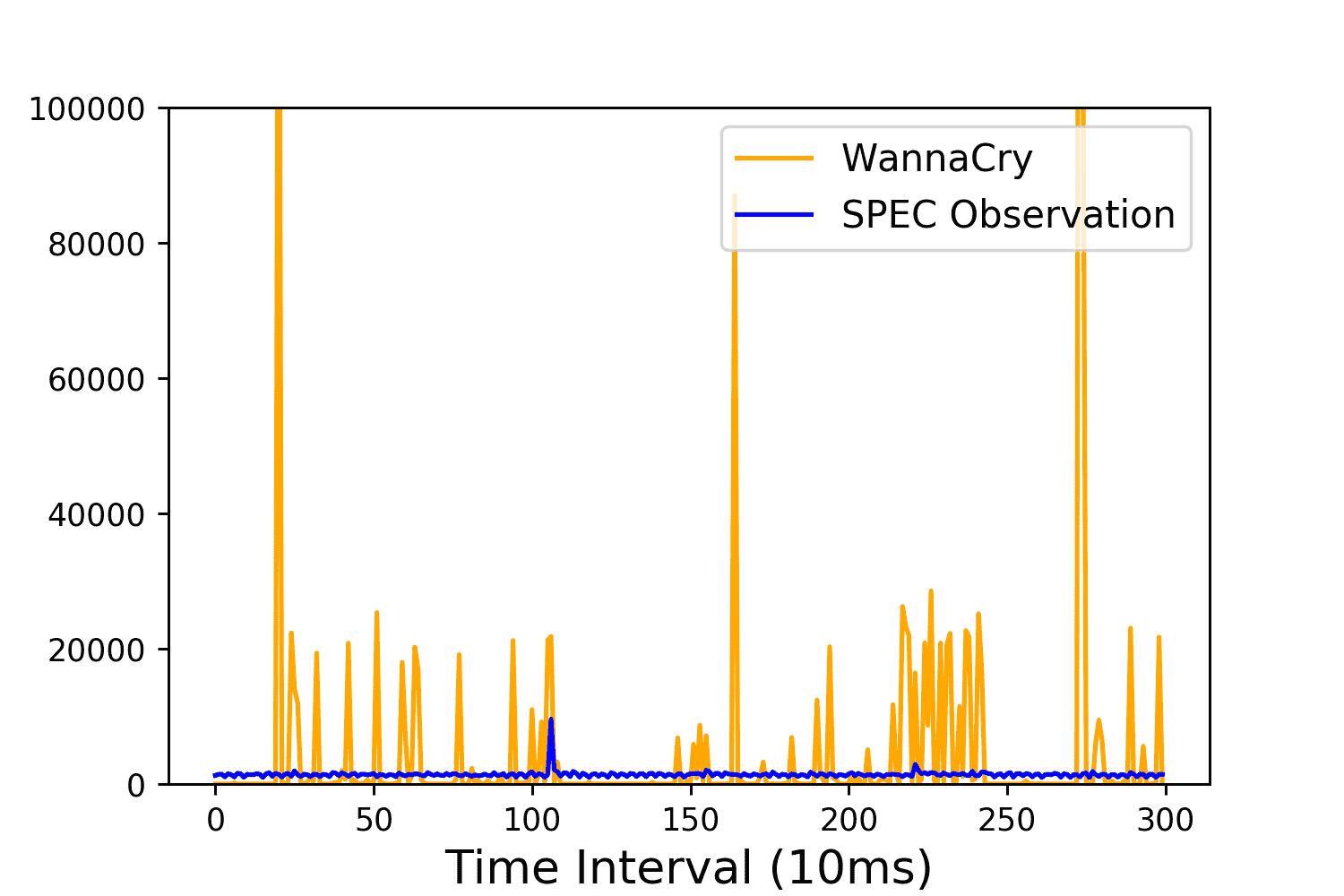}
		\label{fig:com_ca_miss}}
	\subfigure[\textbf{\# Cache References}]{
		\includegraphics[width=0.35\textwidth]{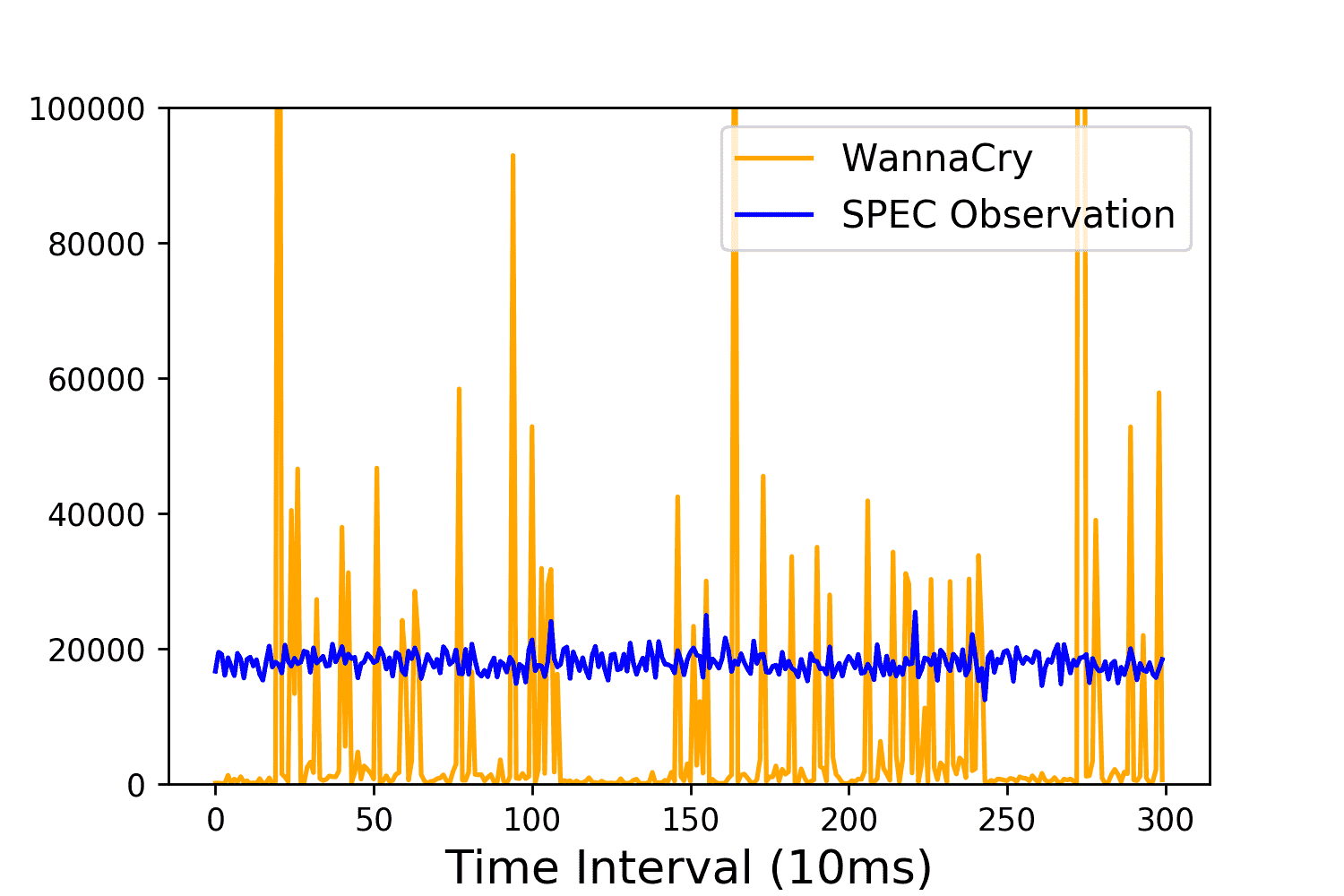}
		\label{fig:com_cache}}
	\subfigure[\textbf{\# Instructions}]{
		\includegraphics[width=0.35\textwidth]{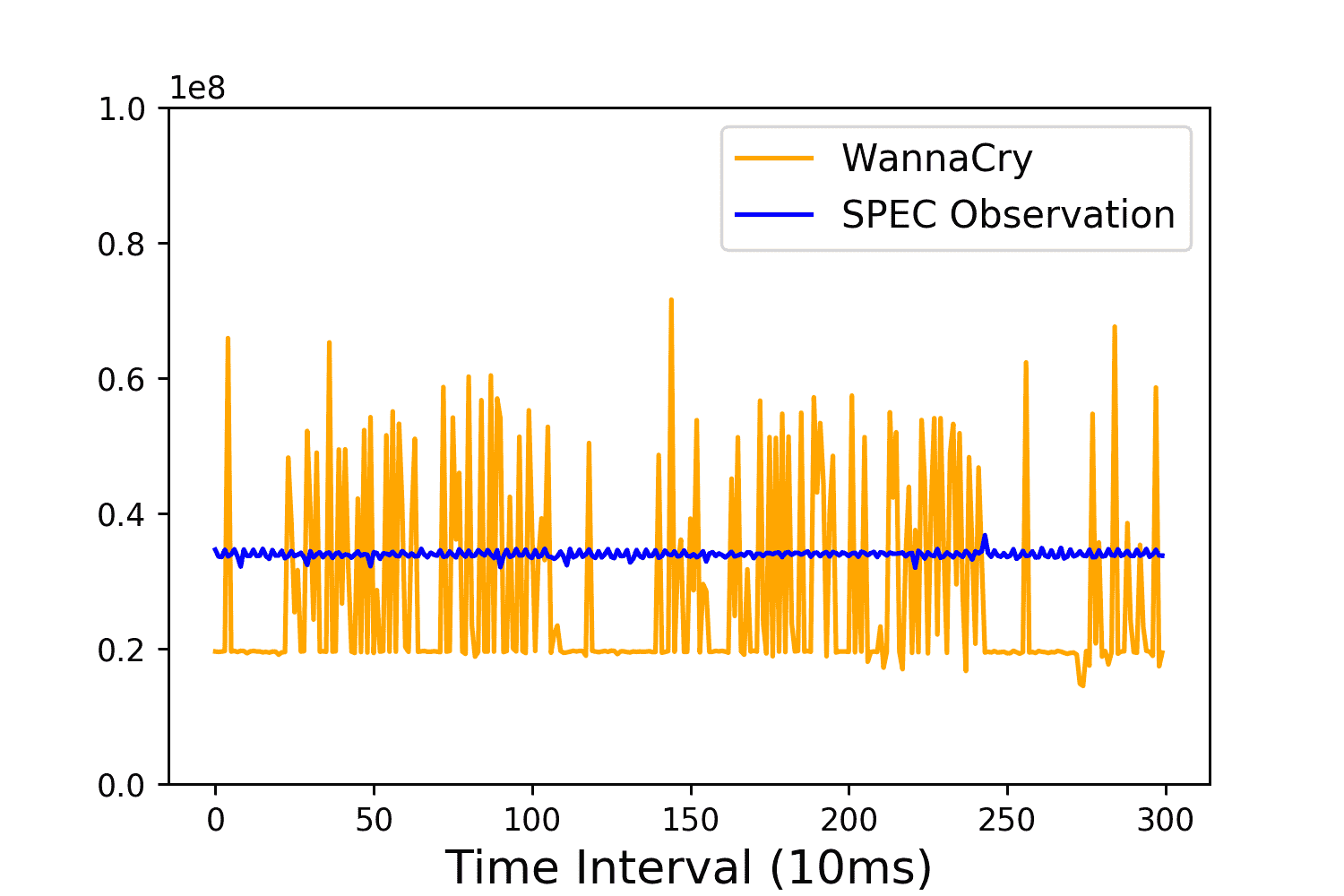}
		\label{fig:com_ins}}
    \end{adjustwidth}
	\vspace{-0.3cm}
	\caption{Comparison of the Effects on Performance Event Counters from HPCs in presence of Wannacry Ransomware and SPEC Benchmark Programs\label{fig:perf_dev_com}\vspace{-0.4cm}}
\end{figure}

\begin{figure}[!b]
	\centering
	\includegraphics[width=0.35\textwidth]{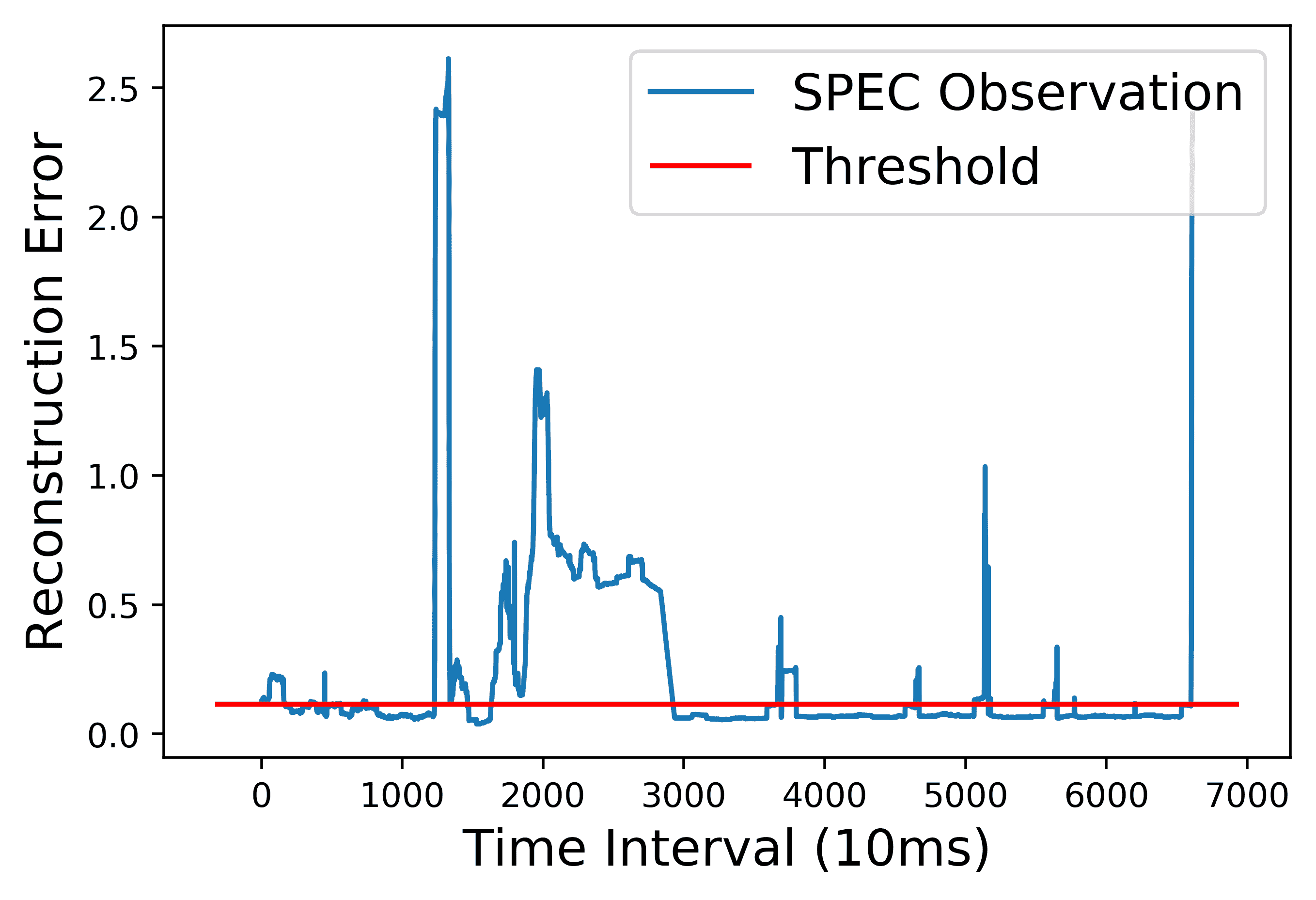}
	\vspace{-0.3cm}
	\caption{Sequence of Reconstruction Errors for SPEC Benchmark in Autoencoder\_1\vspace{-0.5cm}}
	\label{fig:rec_spec}
\end{figure}

In order to test the robustness of our detection scheme, we incorporate an analysis in presence of SPEC2006 server and multimedia benchmarks. We consider the Gshare predictor implementation as provided in $https://www.jilp.org/jwac-2/cbp3\_framework\_instructions.html$ and observe the HPC sampling counts from a background process exactly like our previous setting. Fig.~\ref{fig:perf_dev_com} presents the variation of different hardware events in the presence of both SPEC benchmarks and WannaCry ransomware. We can observe that the execution behaviors for both the programs are significantly different from the normal observations. Thus, the sequences of data for the SPEC programs may also create considerable reconstruction errors. Fig.~\ref{fig:rec_spec} clearly shows that the reconstruction error for the sequences in the presence of the SPEC benchmark programs is above the predetermined threshold at the first window itself. Though the error is very close to the threshold, this essentially raises an alarm to RAPPER that this benchmark program is a potentially malicious program which deviates to an extent from the normal system behavior. But surely, in this case, it is a false alarm, since the benchmark is composed of server and multimedia benchmarks and can be considered as the representative of the high computational processes which may deviate highly from the normal running processes in a system.

In the next subsection, we perform a transformation from the time domain to frequency domain to differentiate actual malicious processes from false positives.

\vspace{-0.3cm}
\subsection{Introducing Fast Fourier Transformation}
\begin{figure}[!t]
	\begin{adjustwidth}{-3cm}{-3cm}
	\centering
	\subfigure[\textbf{Branch Instructions}]{
		\includegraphics[width=0.35\textwidth]{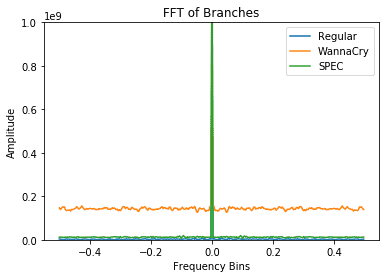}
		\label{fig:fft_branch}}
	\subfigure[\textbf{Branch Misses}]{
		\includegraphics[width=0.35\textwidth]{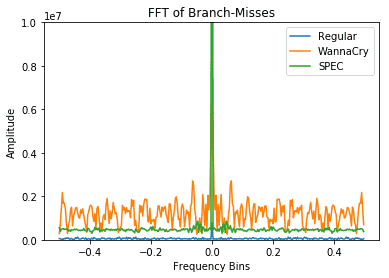}
		\label{fig:fft_br_miss}}
	\subfigure[\textbf{Cache Misses}]{
		\includegraphics[width=0.35\textwidth, height=3.1cm]{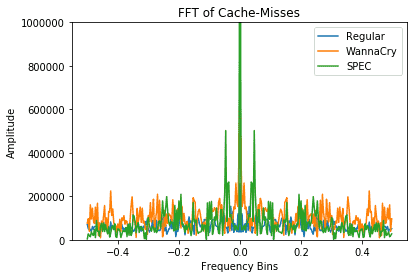}
		\label{fig:fft_ca_miss}}
	\subfigure[\textbf{Cache References}]{
		\includegraphics[width=0.35\textwidth]{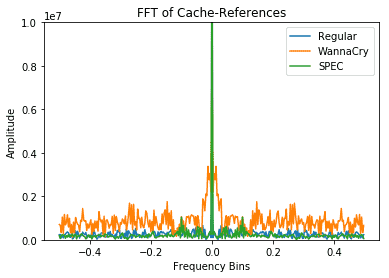}
		\label{fig:fft_cache}}
	\subfigure[\textbf{Instructions}]{
		\includegraphics[width=0.35\textwidth]{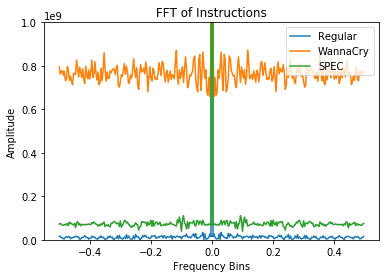}
		\label{fig:fft_ins}}
    \end{adjustwidth}
	\vspace{-0.3cm}
	\caption{Variation of Amplitude in frequency domain of the performance counters from HPCs in the presence of SPEC observation and Wannacry Ransomware\label{fig:fft_dev}\vspace{-0.5cm}}
\end{figure}

In the second phase of detection using RAPPER, we transform the traces from the time domain to the frequency domain using the Fast Fourier Transformation (FFT). FFT is the most efficient way to implement the Discrete Fourier Transformation. The primary reason to convert the analysis from the time domain to the frequency domain is to understand the repetitive pattern of the traces. The ransomware executable runs encryption repeatedly on multiple files thus it repeats the same set of operations of opening a file, encrypting and closing the file followed by deleting it for multiple files one after another. The transformation is illustrated in Fig.~\ref{fig:fft_dev}, which typically indicates that the amplitudes for each frequency bins are constantly higher for the ransomware in contrary to the SPEC benchmark.

We have applied FFT on the time domain values for different hardware events as mentioned in Section~\ref{sec:hpc}, to obtain the frequency domain values. Fig.~\ref{fig:fft_dev} presents the FFT plots for the normal system measurements in \textcolor{blue}{blue} lines, along with the SPEC Observations in \textcolor{OliveGreen}{green} lines and WannaCry Ransomware in \textcolor{orange}{orange} lines for different hardware events. Fig.~\ref{fig:fft_dev} shows that for most of the hardware events (apart from the cache misses), the FFT plot behavior of the SPEC benchmark overlaps exactly with the FFT of the normal system behavior. Also, it is quite clear from Fig.~\ref{fig:fft_branch}, Fig.~\ref{fig:fft_br_miss}, Fig.~\ref{fig:fft_cache}, and Fig.~\ref{fig:fft_ins} that the amplitudes of almost all the frequency bins are higher for WannaCry than the SPEC observation, which is eminent as the WannaCry program repeatedly encrypts multiple files.

\begin{figure}[!t]
	\begin{adjustwidth}{-3cm}{-3cm}
    \centering
    \subfigure[\textbf{SPEC}]{
    	\includegraphics[width=0.35\textwidth]{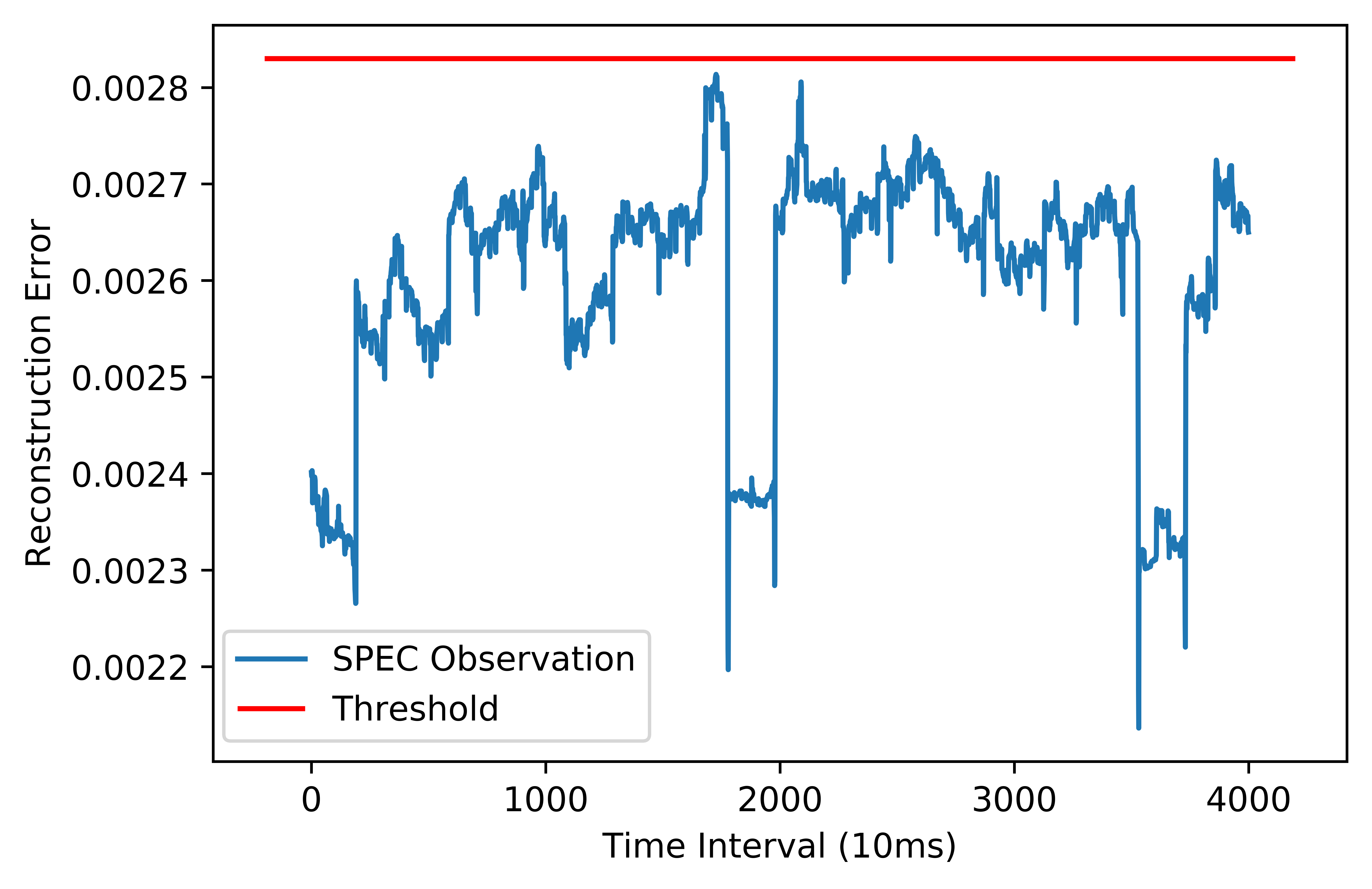}
    	\label{fig:rec_2_spec}}
    \subfigure[\textbf{WannaCry}]{
    	\includegraphics[width=0.35\textwidth]{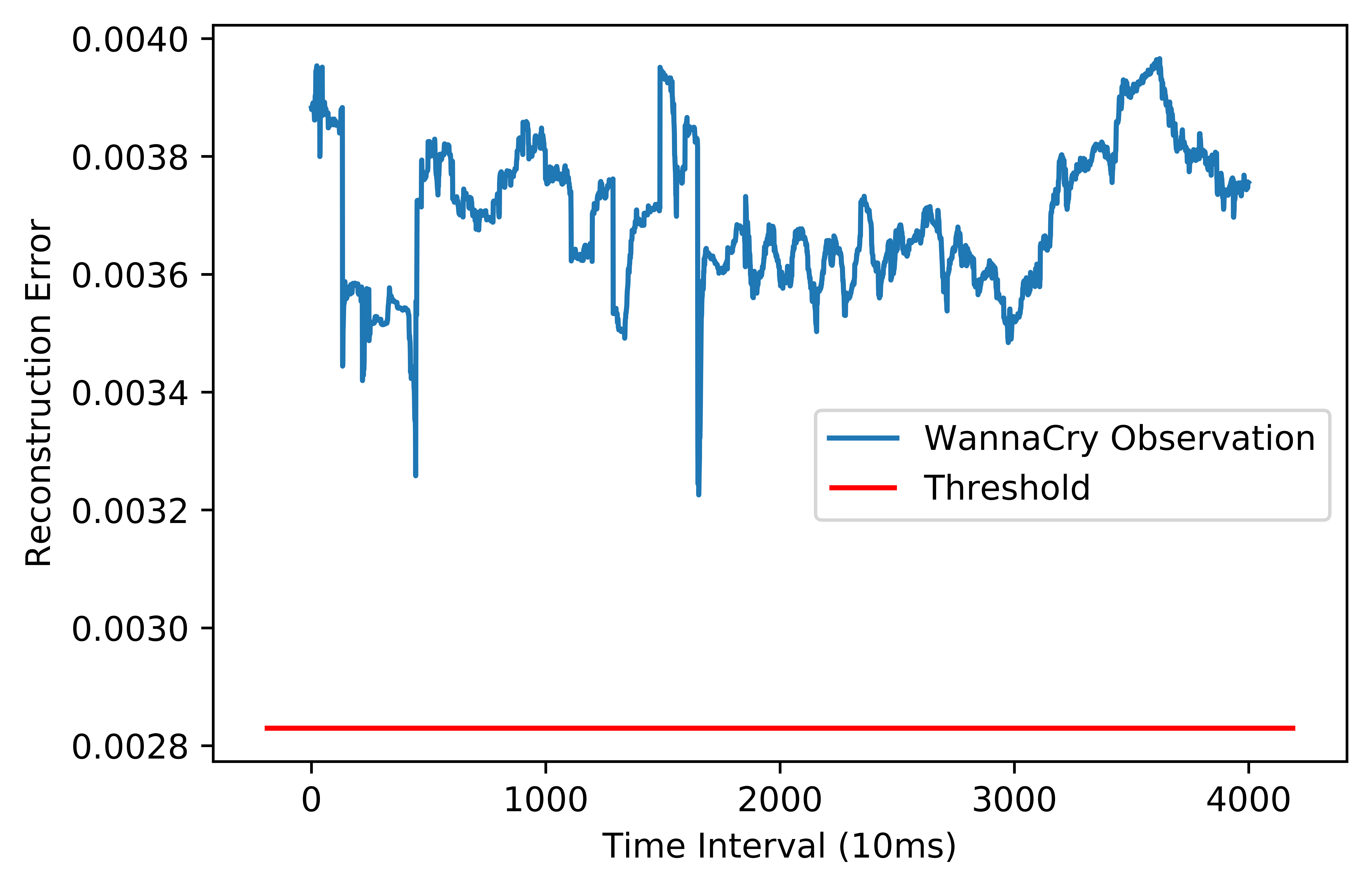}
    	\label{fig:rec_2_wannacry}}
    \subfigure[\textbf{Vipasana}]{
    	\includegraphics[width=0.35\textwidth]{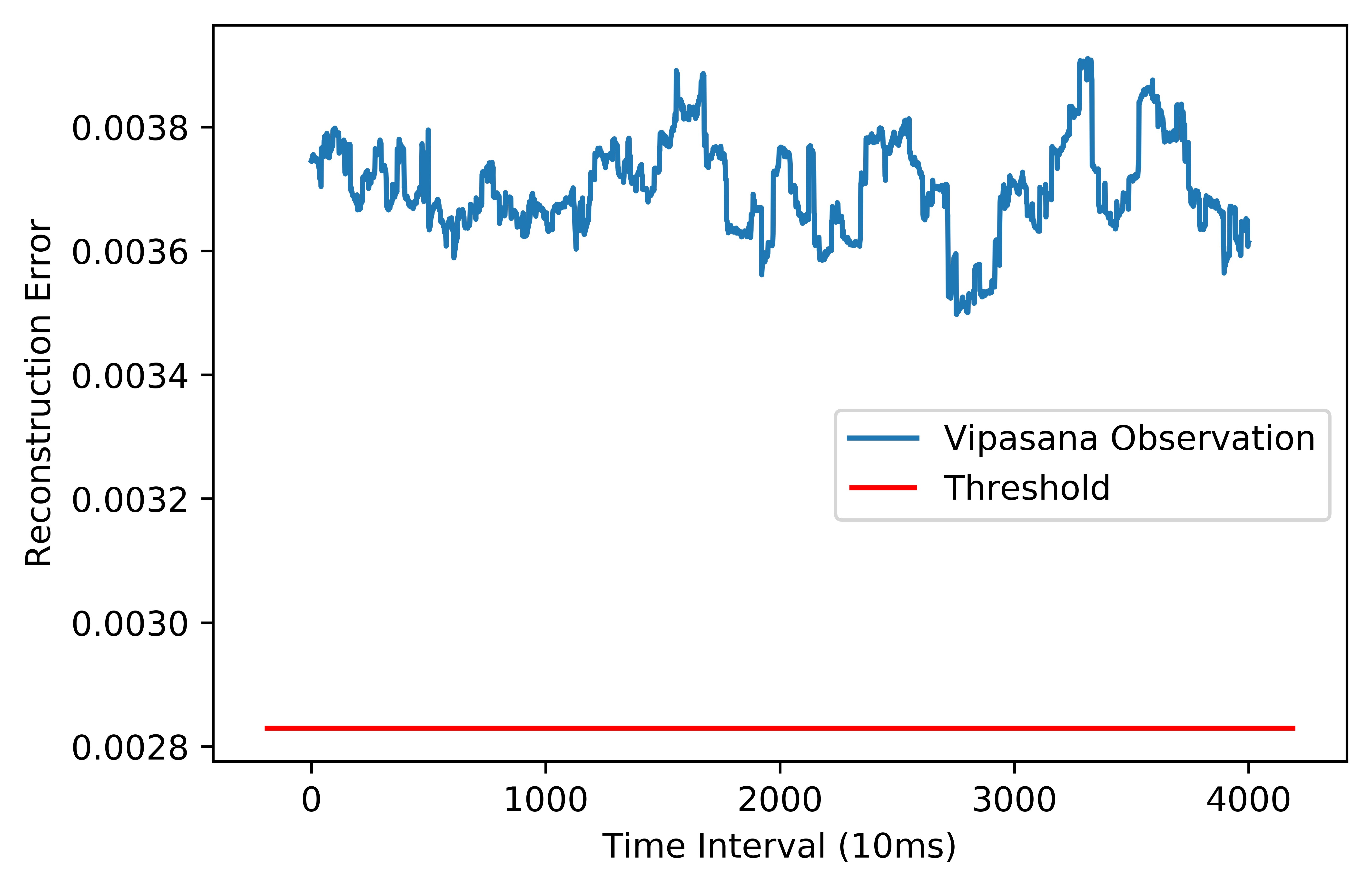}
    	\label{fig:rec_2_vipasana}}
    \subfigure[\textbf{Locky}]{
    	\includegraphics[width=0.35\textwidth]{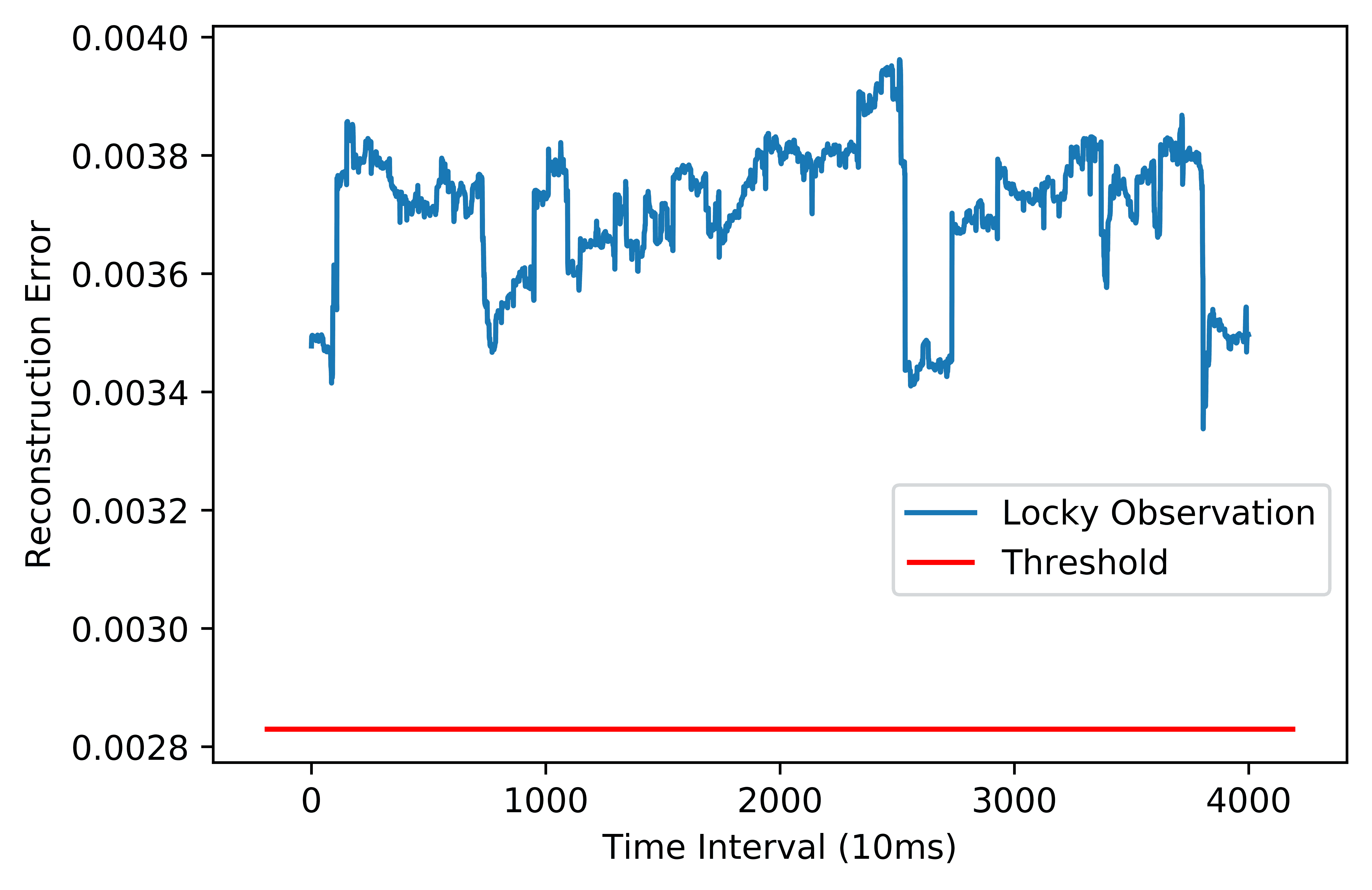}
    	\label{fig:rec_2_locky}}
    \subfigure[\textbf{Petya}]{
    	\includegraphics[width=0.35\textwidth]{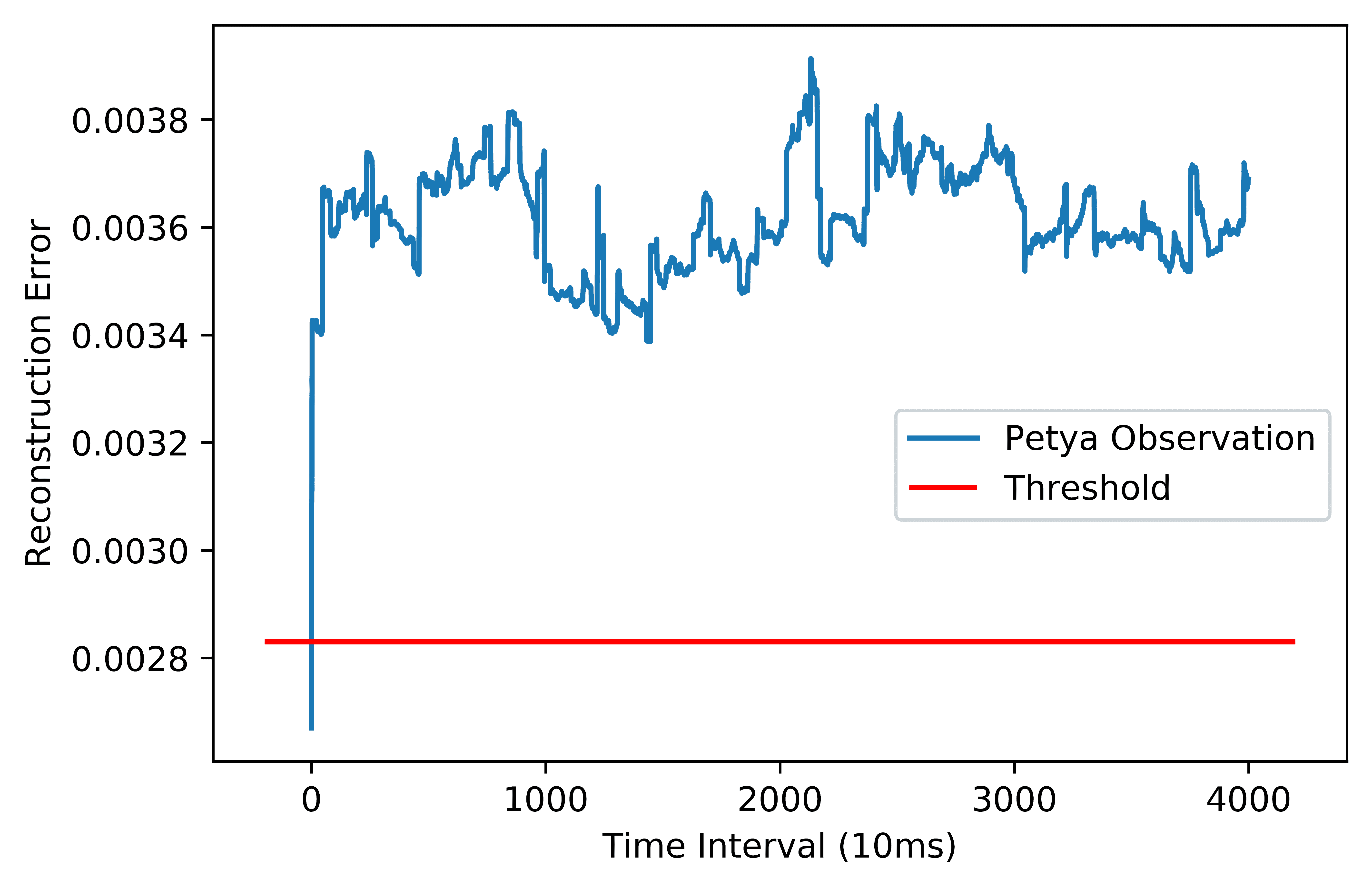}
    	\label{fig:rec_2_petya}}
    \end{adjustwidth}
	\vspace{-0.3cm}
    \caption{Sequence of Reconstruction Errors for SPEC Benchmark and different Ransomwares in Autoencoder\_2\label{fig:seq_rec_error_2}\vspace{-0.5cm}}
\end{figure}

The detection of these variations of amplitudes for different frequency bins can again be considered as a time-series data, and an LSTM based autoencoder, as discussed before, can be used to detect the anomaly. The amplitudes for SPEC benchmark programs are very close to that of regular observations for most of the hardware events. Thus, modeling the FFT data for regular sequences using an autoencoder will result in reconstruction errors close to the threshold (say $\mathcal{R}^{\prime}_t$) for SPEC benchmarks, and the error will be much higher in case of ransomwares because of the repeated encryptions. We modeled another autoencoder following the procedure mentioned before with the FFT transformed data and calculated the threshold $\mathcal{R}^{\prime}_t$ to be $0.002829$. Fig.~\ref{fig:seq_rec_error_2} presents the sequence of reconstruction errors for both SPEC and ransomware programs and we can verify that the reconstruction errors of SPEC programs always lie below the threshold and thus discarded as false positives, whereas the reconstruction errors of all the ransomware programs always remain higher to the threshold\footnote{The reconstruction error for Petya on the first window is lower than the threshold, but for the subsequent windows the error is always higher than the threshold.}.

\vspace{-0.4cm}
\subsubsection{Need of Both the Autoencoders}
One interesting point that may arise in this detection framework is that what is the requirement of both the autoencoders, when it is obvious from the fact that Autoencoder\_2 is sufficient enough to discard the ransomwares. Autoencoder\_2 takes the FFT converted values of the window data, and the FFT conversion requires some time for the computation. Hence, to reduce the computational complexity of the detection framework we apply a first level filter in terms of Autoencoder\_1 to remove the less computational heavy programs in the first stage itself and apply the FFT transformation and Autoencoder\_2 for the anomalous data arising from this stage.

In the next section, we present an analysis on the performance of both the autoencoders in the presence of a standard benign Linux benchmark. 

\vspace{-0.5cm}
\section{Performance on Standard Linux Benchmark}
\vspace{-0.3cm}
\begin{figure}[!t]
	\centering
	\subfigure[\textbf{Autoencoder\_1}]{
		\includegraphics[width=0.35\textwidth]{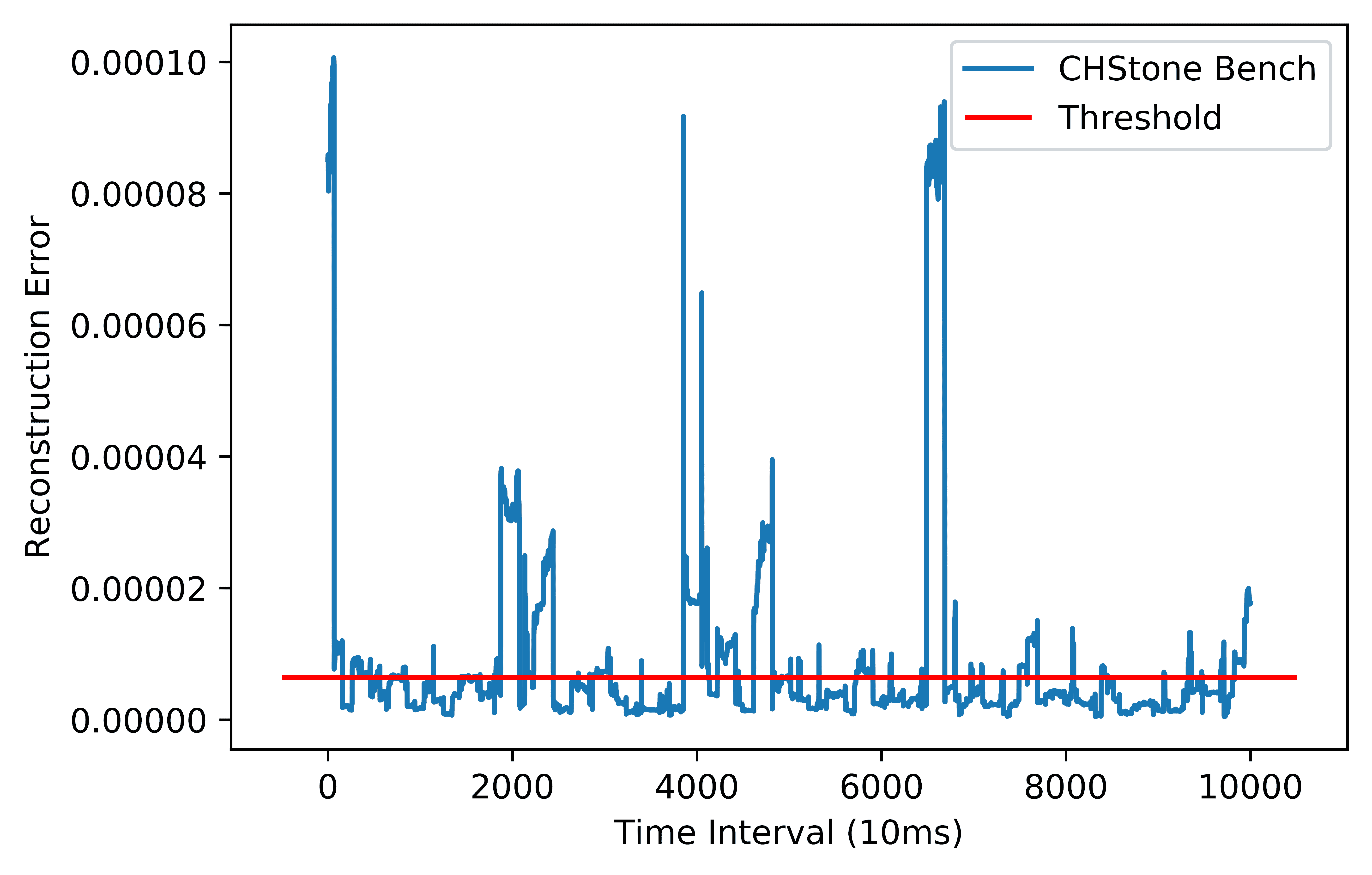}
		\label{fig:chstone_1}}
	\subfigure[\textbf{Autoencoder\_2}]{
		\includegraphics[width=0.35\textwidth]{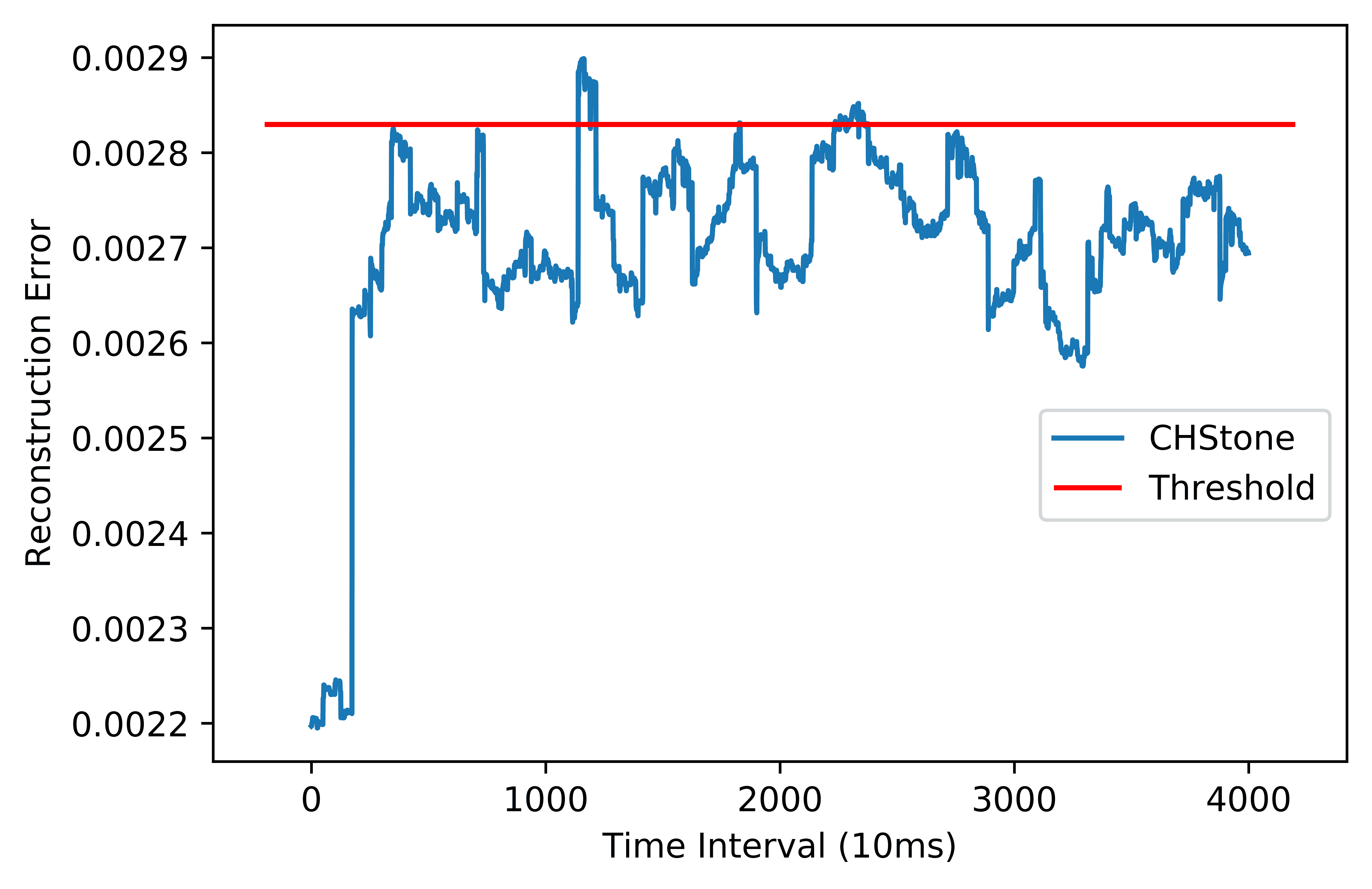}
		\label{fig:chstone_2}}
	\vspace*{-0.3cm}
	\caption{Sequence of Reconstruction Errors on the CHStone benchmark.\label{fig:chstone}\vspace*{-0.5cm}}
\end{figure}

In this section, we consider a standard Linux benchmark CHStone~\cite{chstone:2008} to analyze the efficiency of both the autoencoders. CHStone is Linux benchmark suite which represents various application domains such as arithmetic, media processing, and security. Hence, it would be intriguing to evaluate the performance of the detection scheme in the presence of this benchmark suite. The sequence of reconstruction errors for both the autoencoders in this scenario is shown in Fig.~\ref{fig:chstone}. We can easily observe from Fig.~\ref{fig:chstone_1} that in most of the cases the error value in the first autoencoder is lower than the threshold. However, in some of the cases, the error is higher (i.e., it is detected as an anomaly), but the reconstruction error, as shown in Fig.~\ref{fig:chstone_2}, in the second autoencoder is always lower than the threshold (except at some specific time interval). We can hypothesize from this example that, if the reconstruction error in both the autoencoders are constantly higher for some specific time we conclude that behavior as anomaly instead of considering a single peak. We have also experimented with two other benchmarks such as Unix-Bench~\cite{unix_bench} and LMBench~\cite{lmbench}, and the results are similar in nature.

\vspace{-0.5cm}
\section{Identifying Disk Encryption from Ransomwares}
\vspace{-0.3cm}
\begin{figure}[!t]
    \centering
    \subfigure[\textbf{}]{
    	\includegraphics[width=0.35\textwidth]{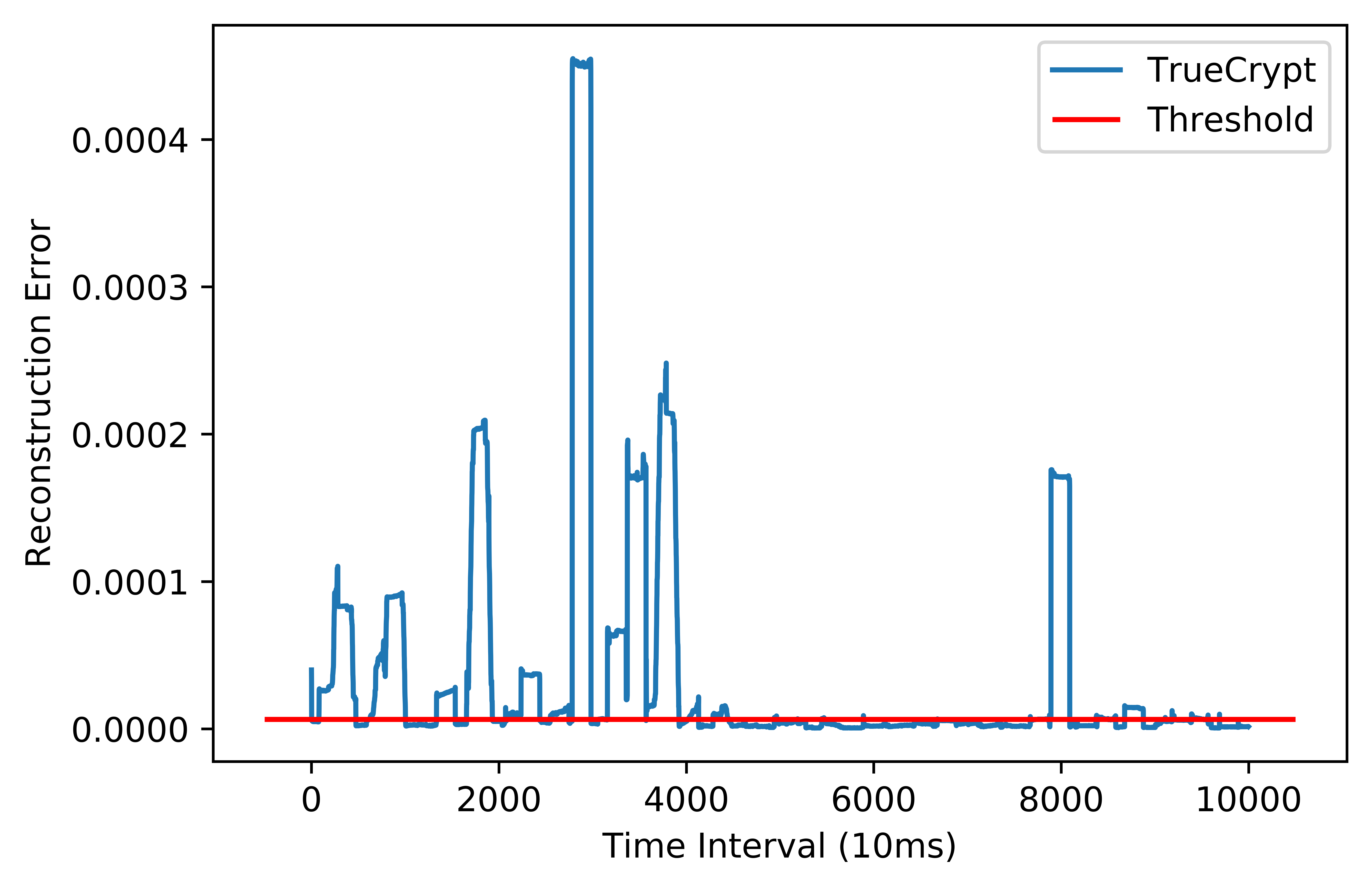}
    	\label{fig:rec_disk_true_1}}
    \subfigure[\textbf{}]{
    	\includegraphics[width=0.35\textwidth]{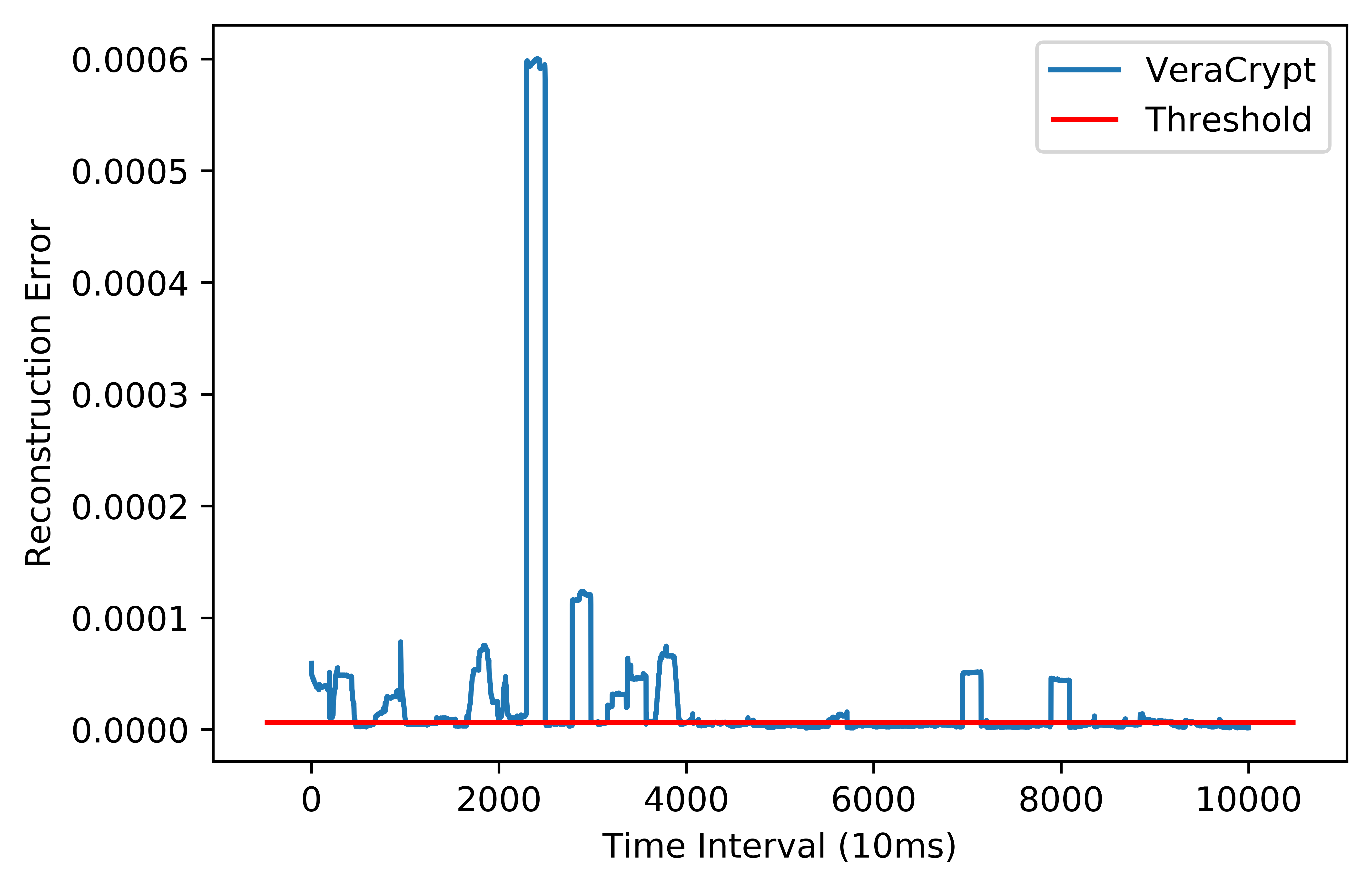}
    	\label{fig:rec_disk_vera_1}}
    \subfigure[\textbf{}]{
    	\includegraphics[width=0.35\textwidth]{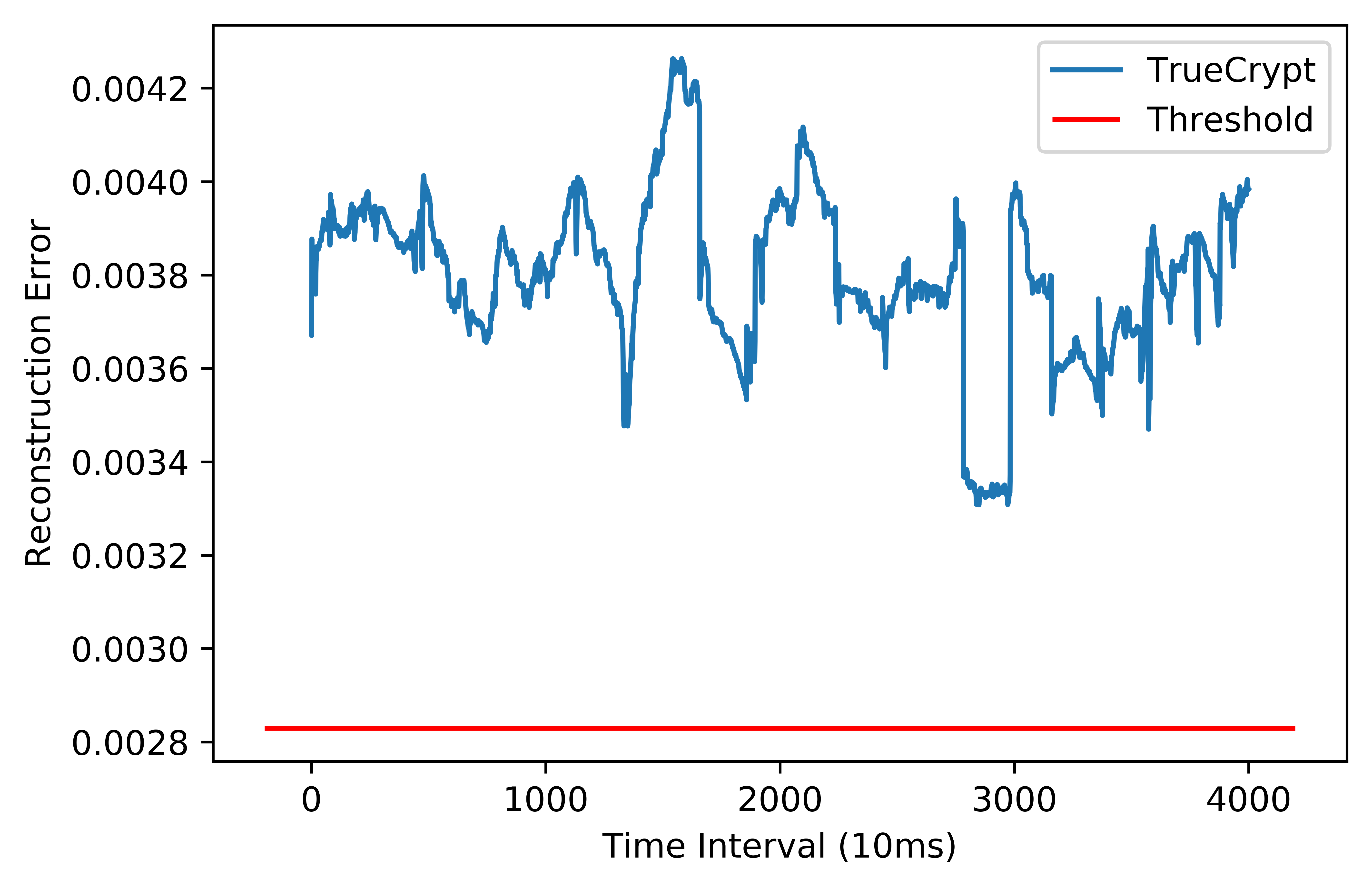}
    	\label{fig:rec_disk_true_2}}
    \subfigure[\textbf{}]{
    	\includegraphics[width=0.35\textwidth]{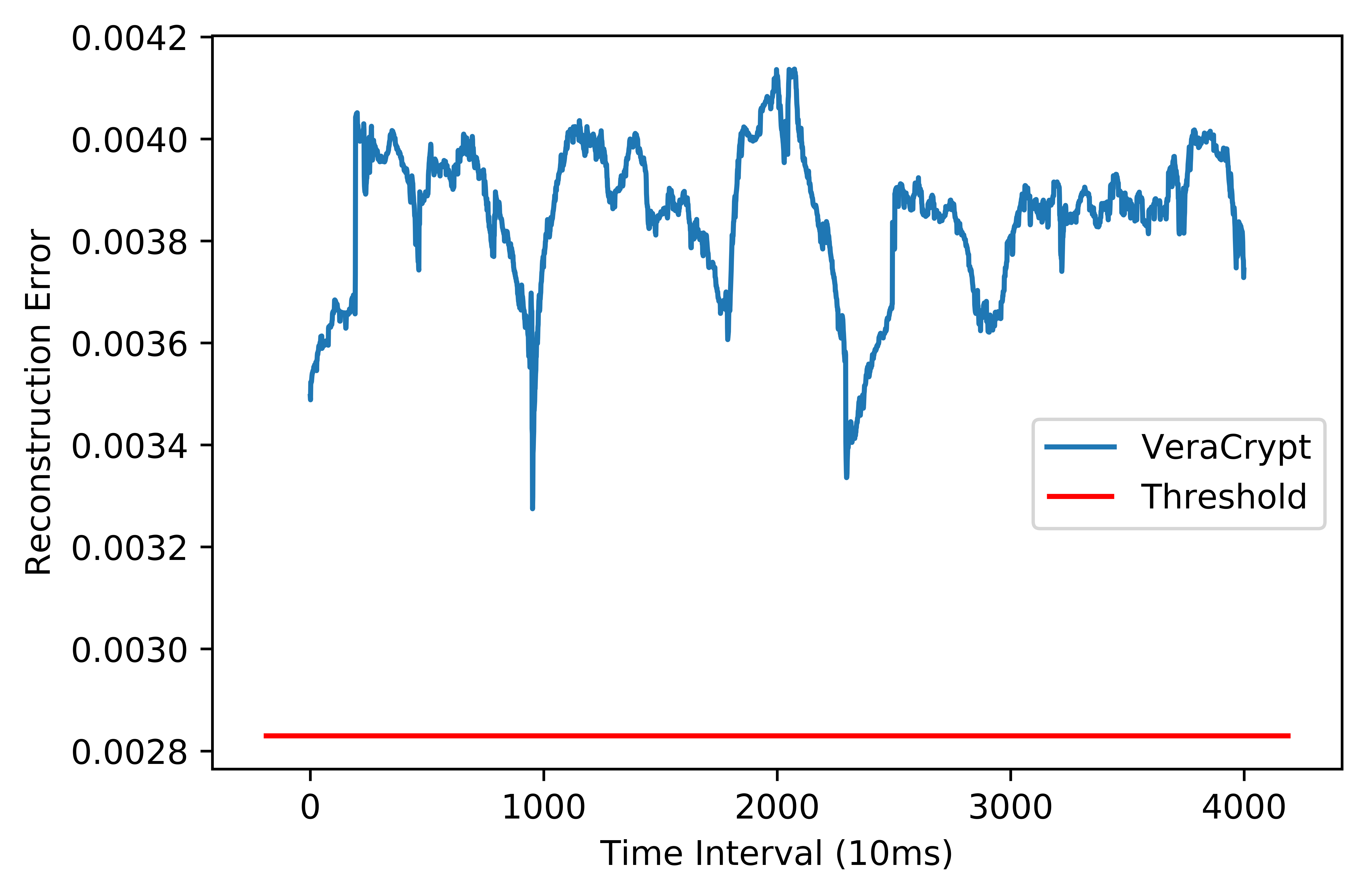}
    	\label{fig:rec_disk_vera_2}}
    \vspace{-0.3cm}
    \caption{Sequence of Reconstruction Errors for Disk Encryption Programs in both the Autoencoders. \textbf{(a)} TrueCrypt in Autoencoder\_1, \textbf{(b)} VeraCrypt in Autoencoder\_1, \textbf{(c)} TrueCrypt in Autoencoder\_2, \textbf{(d)} VeraCrypt in Autoencoder\_2\label{fig:seq_rec_error_disk}\vspace{-0.5cm}}
\end{figure}

The Disk Encryption processes are very similar in operation to the malicious ransomware processes. Both of these processes access files frequently and encrypt them one after another, though the intentions of the processes are entirely different. While designing RAPPER, one of the most significant challenges is to differentiate the disk encryption processes from malicious ones. Also, there are some ransomwares which uses disk encryption program as their encryption engine. This typically puts the security engineers in a very delicate state. In our paper, we will be discussing how the disk encryption programs differs from the ransomware processes which does not use disk encryption as their intermediate software routine. This detection module helps us to differentiate the disk encryption modules from the general set of ransomware programs. Later we use this detection module to demonstrate a reasonable solution to this problem of ransomware detection.

In order to manifest the problem, we consider two popular disk encryption processes in our study, namely \emph{TrueCrypt}, and \emph{VeraCrypt}. The behavior of both the processes in both the autoencoders are shown in Fig.~\ref{fig:seq_rec_error_disk}. We can easily see from the figure that, both the disk encryption processes are detected as ransomwares by RAPPER. \emph{One naive solution is to check the privilege of the current process under suspicion}. Since the disk encryption processes can only be run by the administrator with the highest privilege, checking the privilege of the running application can be a quick check to determine whether the target process is malicious or not. 

In this paper, however, we have also delved into the harder problem of differentiating these two sets of processes by looking at the nature of the HPC event values. All the popular disk encryption processes use AES-XTS\footnote{AES XEX-based tweaked-codebook mode with ciphertext stealing.} mode of encryption for their operations. We utilize this characteristic to template the operation of a disk encryption process, and in the online phase, we check for whether the suspicious program is a disk encryption process or not. In order to find similarity with the stored template, we calculate the cumulative correlation of it with the suspicious process. If the correlation is high for a successive interval of time, we conclude that the process is a disk encryption process and allow it to execute in the system.

The watchdog program generates a successive window of multivariate data with significantly different amplitude values for various applications. Instead of using complex multivariate correlation, we use the univariate reconstruction error from the Autoencoder\_1 for the simple \emph{Pearson's correlation} to make the detection less computationally expensive and with less storage requirement. We store the template of reconstruction errors for a disk encryption process instead of multivariate window data and correlate it with the reconstruction errors of the unknown process.

\begin{figure}[!t]
	\centering    \includegraphics[width=8cm, height=3cm]{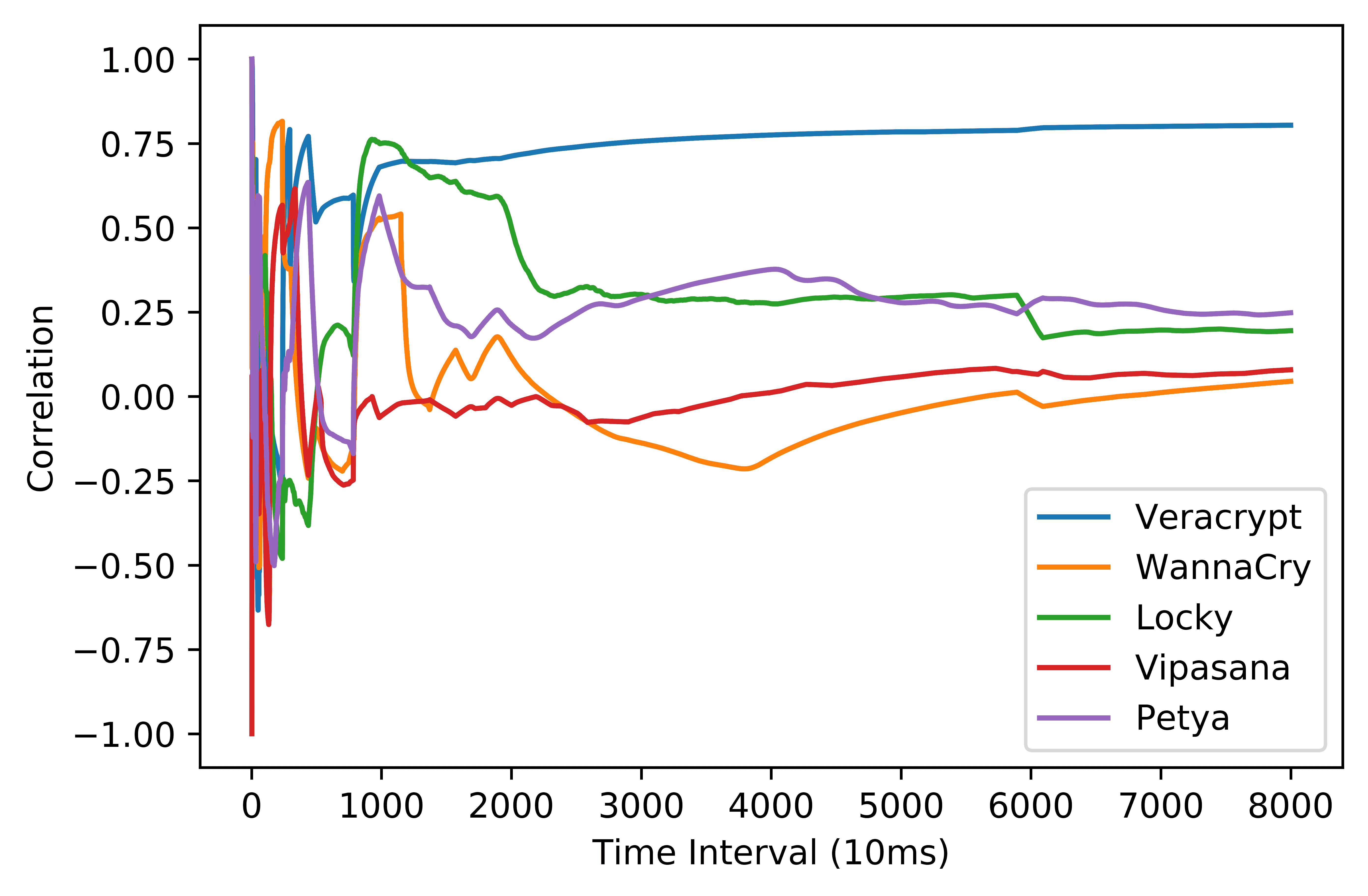}
	\vspace{-0.3cm}
    \caption{Cumulative Correlation of Veracrypt disk encryption and Ransomware processes with another disk encryption process TrueCrypt\label{fig:correlation}\vspace{-0.5cm}}
\end{figure}

In order to demonstrate the approach we use reconstruction errors of \emph{TrueCrypt} as our template, and present the cumulative correlation values with \emph{VeraCrypt} and other ransomwares in Fig.~\ref{fig:correlation}. We can easily observe that the correlation values of VeraCrypt are high for a successive interval of time, whereas, the correlation values for all the ransomwares converges to a very low value. Hence, we conclude that it is easy to differentiate the behavior of disk encryption programs from ransomwares with the hypothesis that most of the popular disk encryption programs use the same mode of encryptions in their operations.

\vspace{-0.5cm}
\subsubsection{Comprehensive detection and temporary suspension of disk Encryption Processes}
The discussion in the previous section shows that a particular mode of encryption can be differentiated with high confidence if the HPC events are monitored in an efficient manner. This identification specifically means that all disk encryption algorithms running AES in XTS mode can be differentiated from the general genre of malicious ransomware programs which have no disk encryption sub-routine in them. Though as mentioned earlier, their exists some ransomware like MAMBA which uses disk encryption modules so as to maliciously encrypt process. Our detection module as described in the previous subsection can  successfully identify that whether a disk encryption module is running in the background but it turns out that the disk encryption could also be a part of ransomware operation. In this paper, we propose a solution to this problem by temporary suspending the suspected disk encryption program. This temporary suspension of the disk encryption program raises an alarm to the user and waits for a confirmation from the user whether the suspected program is actually launched by the user. This confirmation will automatically resume the disk encryption module intended to run from the user's end but not the unintended ones which gets launched by the ransomware modules.

In the next section, we discuss in details the overall detection strategy of RAPPER with a flow diagram.

\vspace{-0.5cm}
\section{Architecture of RAPPER}
\vspace{-0.3cm}
In this section, we present an overview of the architecture of proposed detection methodology - RAPPER. The basic diagram of the system is shown in Fig.~\ref{fig:architecture}. All the experimentation for this study have been performed in a sandbox environment, such that the ransomwares do not affect the actual file system. The architecture contains five modules: \emph{Watchdog Program}, \emph{Autoencoder\_1}, \emph{FFT Converter}, \emph{Autoencoder\_2}, and \emph{Correlation Module}. The detection methodology works in two phases, namely \emph{Offline Phase} and \emph{Online Phase}. The functioning of each module in both the phases are described below:

\begin{figure}[!t]
	\centering
	\includegraphics[width=1\textwidth, height=6cm]{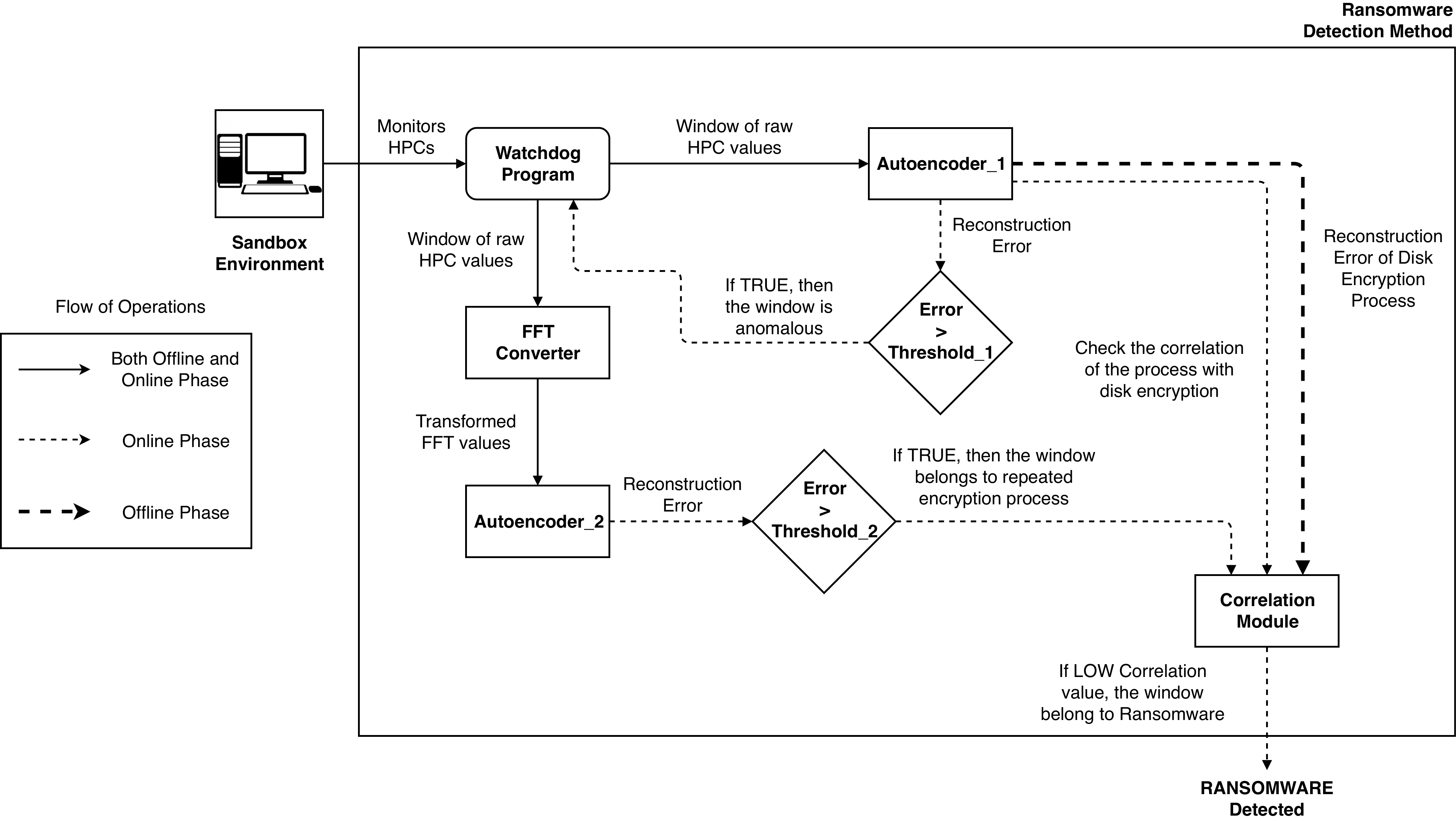}
	\vspace*{-0.75cm}
	\caption{Detection Methodology of RAPPER\label{fig:architecture}\vspace*{-1cm}}
\end{figure}

\vspace{-0.5cm}
\subsection{Offline Phase}
\vspace{-0.3cm}
In the offline phase, the detection methodology is trained with the normal behavior of the sandbox environment, such that any unusual activity of a ransomware is properly detected in real-time scenario. The functioning of each of the modules in this phase are described below.
{\small 
\begin{enumerate}
\item \emph{Watchdog Program}: Monitors the HPCs of the Sandbox Environment continuously and forwards a window of data to the Autoencoder\_1 and the FFT Converter in parallel. 
\item \emph{Autoencoder\_1}: Trains an autoencoder with the dataset forwarded by watchdog program. This module also forwards reconstruction error corresponding to a disk encryption process to the correlation module directly\footnote{Behavior of disk encryption is not included in the training as this may produce false negatives for ransomwares since both are repeated encryption process. The errors due to disk encryption are calculated after the training is completed.}.
\item \emph{FFT Converter}: Computes Fast Fourier Transformation of each window forwarded by watchdog program and passes the results to the Autoencoder\_2.
\item \emph{Autoencoder\_2}: Collects all the data passed by the FFT Converter and trains another autoencoder based on the FFT dataset.
\item \emph{Correlation Module}: Stores reconstruction errors corresponding to a disk encryption for analysis in the \emph{Online Phase}.
\end{enumerate}}

\vspace{-0.5cm}
\subsection{Online Phase}
\vspace{-0.3cm}
In the online phase, the detection module is deployed in the sandbox system for real-time monitoring to detect ransomwares. The functioning of each modules for an unknown process in this phase are discussed below.
{\small
\begin{enumerate}
\item \emph{Watchdog Program}: Monitors the system and forwards data to the Autoencoder\_1. In this phase, watchdog program does not forward data to the FFT converter for run-time monitoring the system with lower computational cost. 
\item \emph{Autoencoder\_1}: Calculates reconstruction error of the data received from watchdog program. If the error is higher than the predefined threshold $\mathcal{R}_t$, it sends a signal to the watchdog program to transmit the same window to the FFT Converter. Otherwise, the process is allowed to execute in the system. This module also forwards the data to correlation module directly, irrespective of it being lesser or more than the threshold value. 
\item \emph{FFT Converter}: Converts the data received from watchdog program into frequency domain, and forwards the transformed data to the Autoencoder\_2, but with a condition imposed by the Autoencoder\_1 module.
\item \emph{Autoencoder\_2}: Calculates the reconstruction error of the received FFT data, and if the error is higher than the predefined threshold, $\mathcal{R}^{\prime}_t$, it sends a signal to the correlation module to check for its correlation with the template of disk encryption process. Otherwise, the process is considered as simply a high computational process and is allowed to execute in the system. 
\item \emph{Correlation Module}: Calculates the cumulative correlation of the unknown process with the known disk encryption process. If the correlation is low for a considerable duration of time, then the process is considered as a Ransomware and is terminated from the system, else it is forwarded to the user for verification as a legitimate disk encryption process.
\end{enumerate}}

\vspace{-0.3cm}
In the next section, we analyze the efficiency of RAPPER in terms of detection time of ransomwares and implementation overhead.

\vspace{-0.5cm}
\section{Evaluating the performance of RAPPER}\label{sec:results}
\vspace{-0.3cm}

We performed all the experiments in a sandbox system having specification \texttt{Intel Core i3 M350} running Linux 4.10.0-38-generic kernel. We used popular open source python based neural network library \texttt{Keras}~\cite{keras} for the implementation of both the autoencoders. 




The FFT converter usually takes $0.0003$ milliseconds to convert a sequence within a window into frequency domain. The model building times for Autoencoder\_1 and Autoencoder\_2 are on average $10$ and $14$ minutes respectively. Testing time to calculate whether a single window is an anomaly or not is $1.321$ milliseconds for Autoencoder\_1 and $1.699$ milliseconds for Autoencoder\_2 respectively. As shown in the Architecture of RAPPER in Fig.~\ref{fig:architecture}, the testing of a regular observation only passes through the Autoencoder\_1, thereby taking only $1.699$ milliseconds, and an anomalous observation passes through all the three modules: Autoencoder\_1, FFT Converter, and Autoencoder\_2, thereby taking $1.321+0.0003+1.699 = 3.0203$ milliseconds to be detected. The time to correlate two reconstruction errors from Autoencoder\_1 and the stored error trace is on an average $0.0001$ milliseconds, which will be calculated only for either disk encryption or ransomwares. In both the cases of regular and anomalous observation, the detection time is less than the sampling interval, which is $10$ milliseconds. Hence, \emph{the detection is performed seamlessly, without the need of any storage buffer}, as a new window of data will be created after 10 milliseconds.

Without loss of generality, we present here the calculation of detection time for most recent WannaCry ransomware by RAPPER. As shown in Section~\ref{sec:threshold}, the WannaCry is detected as an anomaly at the $432^{nd}$ window and instantly detected as ransomware at the same time because it's reconstruction error is always higher than the threshold of Autoencoder\_2. Hence, the total time taken to detect WannaCry as a repeated encryption process is equal to (Time taken to generate the first window) + $431$ * (time interval for each sample) + (Autoencoder\_1 testing time) + (Time for single FFT Conversion) + (Autoencoder\_2 testing time) $=1000+431*10+1.321+0.0003+1.699$ millisecond $=5313.0203$ milliseconds. Thus, WannaCry is detected by RAPPER as a repeated encryption process in approximately $5.313$ seconds. We can check the privilege of the program and terminate it instantly as ransomware does not have administrator privilege. However, the correlation module discriminates it from the disk encryption process from $1002^{nd}$ window, as shown in Fig.~\ref{fig:correlation}. So, for confirming it as a ransomware program, by correlation, it takes almost extra 5 seconds of execution time. Since, at this stage we check for a suspicious ransomware process, we can always pause the execution of such process and resume it after the verification. As a sample run with RAPPER, out of 10000 files of approximately 21 bytes each, when the detection stops the execution, 68 files are encrypted. It maybe noted that, the size of a typical file is much larger than 21 bytes, and hence, a lesser number of files will be encrypted.


\vspace{-0.5cm}
\section{File Recovery and Conclusions}
\vspace{-0.3cm}
\begin{figure}[!t]
	\centering
    \includegraphics[width=0.75\textwidth]{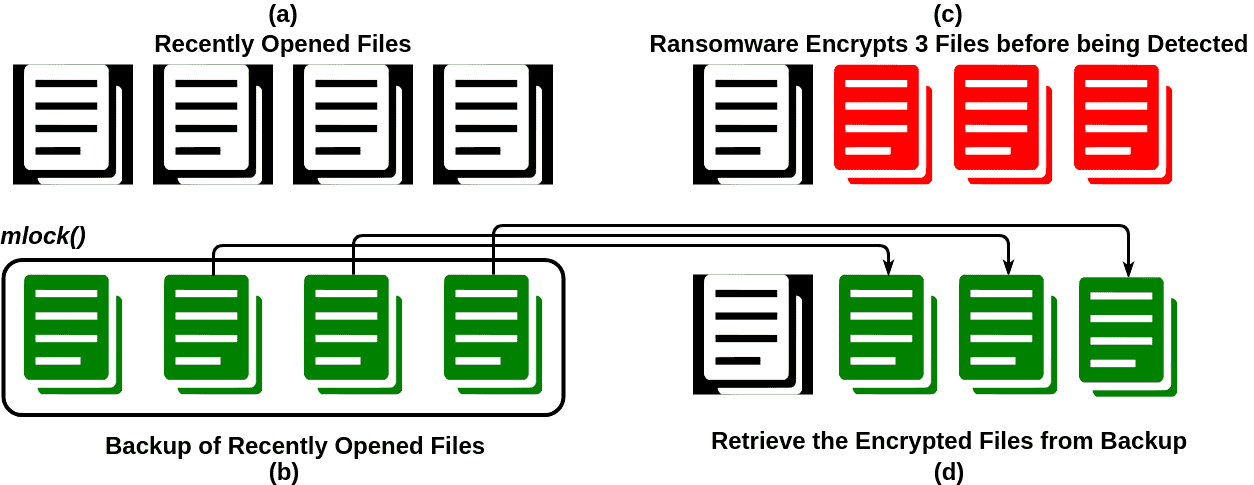}
    \vspace{-0.3cm}
    \caption{Notion of File Recovery using Linux \texttt{mlock()}. \textbf{(a)} Let there are 4 files which are opened within a specific time quantum. \textbf{(b)} Backup these files with Linux \texttt{mlock()} command (marked with \textcolor{OliveGreen}{green} color). \textbf{(c)} Let ransomware encrypts 3 files before being detected by RAPPER (marked with \textcolor{red}{red} color). \textbf{(d)} We can easily retrieve the encrypted files from the backup.\label{fig:file_recover}\vspace{-0.2cm}}
\end{figure}

RAPPER is thus capable of detecting the presence of ransomwares fast, as we show for the case of WannaCry within a time of approximately 5 sec from its launch. Depending on the latency, the malware can encrypt a few files (say $n$). We conclude with a suggested approach for data retrieval. A practical solution would be to take backups of the $n$-recently opened files. After the lapse of the time quantum required to encrypt these files, we delete the copies if RAPPER raises no ransomware alarm. This minimizes the storage requirement for the backup files. To further ensure that the backup files are not encrypted we perform locking operation, like in Linux using {\tt mlock()}. A basic idea of this approach is presented in Fig.~\ref{fig:file_recover}, which shows the operation of Linux \texttt{mlock()} in order to recover the encrypted files in presence of the ransomware. 

In this paper, we  provide a detailed understanding on the effect of ransomware on normal system behaviors. We take the aid of the Artificial Neural Network to detect the presence of ransomwares using a two-step detection framework. The entire detection procedure does not need any template of the malicious process from beforehand. Instead it thrives on an anomaly detection procedure to detect the infectious ransomwares in as less as 5 seconds with almost zero false positives, using a frequency analysis. 

We also explored the opportunity of applying side channel techniques to recover the secret key used to encrypt the files from the perf statistics. We found for ransomwares like WannaCry; each file is encrypted using AES-128 CBC (Cipher Block Chaining) with a randomly generated distinct key. These keys are in turn encrypted using an infection specific RSA public key and stored in the memory. It would be indeed a challenging exercise to recover the AES key by targeting the AES CBC operation. However, we leave that as a future scope of work.

\bibliographystyle{splncs03}
\bibliography{refs}

\end{document}